\documentclass[twocolumn,twocolappendix]{aastex631}

\usepackage{nccmath}
\usepackage{mathtools}
\newcommand{\Tb}{Table }
\newcommand{\Fg}{Figure }
\newcommand{\Eq}{Eq.}
\newcommand{\Sec}{Section }
\newcommand{\Ap}{Appendix }

\begin{document}

\title{High-energy neutrinos from late-time jets of gamma-ray bursts seeded with cocoon photons}
\correspondingauthor{Riki Matsui}
\email{riki.matsui@astr.tohoku.ac.jp}

\author[0000-0003-0805-7741]{Riki Matsui}
\affiliation{Astronomical Institute, Graduate School of Science, Tohoku University, Sendai 980-8578, Japan}

\author[0000-0003-2579-7266]{Shigeo S. Kimura}
\affiliation{Frontier Research Institute for Interdisciplinary Sciences, Tohoku University, Sendai 980-8578, Japan}
\affiliation{Astronomical Institute, Graduate School of Science, Tohoku University, Sendai 980-8578, Japan}

\author[0000-0003-2866-4522]{Hamid Hamidani}
\affiliation{Astronomical Institute, Graduate School of Science, Tohoku University, Sendai 980-8578, Japan}

\begin{abstract}
In gamma-ray bursts (GRBs), $\sim$ 100 - 1000 s after the prompt emission, afterglow observations have consistently shown X-ray excesses detected in the form of flares (XFs; in long GRBs) or extended emission (EEs; in short GRBs).
These observations are interpreted as emissions from jets launched by late central engine activity. 
However, the characteristics of these late-time jets, particularly the dissipation radius ($r_{\rm diss}$), Lorentz factor ($\Gamma$), and cosmic-ray loading factor ($\xi_p$), remain unknown despite their importance.
Here, in order to understand the properties of the late-time jets with future multi-messenger observations, we estimate the detectability of neutrinos associated with late-time emissions for a wide range of $r_{\rm diss}$ and $\Gamma$, assuming $\xi_p=10$.
We take into account external seed photons from the cocoon around the jets, which can enhance the neutrino production through photohadronic interaction in the jet dissipation region.
Our results are still consistent with the upper limit obtained by IceCube.
Our calculations indicate a promising prospect for neutrino detection with IceCube-Gen2 through the stacking of $\sim 1000-2000$ events, for a wide range of $r_{\rm diss}$ and $\Gamma$.
We found that setting an optimal energy threshold of 10 TeV can significantly reduce noise without negatively affecting neutrino detection. 
Furthermore, even in the case of non-detection, we show that meaningful constraints on the characteristics of the late-time jets can be obtained.
\end{abstract}

\keywords{Neutrino astronomy (1100) --- Particle astrophysics (96) --- Gamma-ray bursts (629) --- Gravitational wave astronomy(675) }

\section{Introduction} \label{sec:intro}

Gamma-ray bursts (GRBs) are produced by relativistic jets \citep[e.g.][]{Schmidt1978Natur.271..525S,Paczynski1986,Goodman1986ApJ...308L..47G}.
 GRB-jets are expected to power two types of emissions: the prompt emission and the afterglow emission.
The prompt emission is the early component characterized by intense gamma-ray emissions lasting 0.01 to 100 s \citep{Kouveliotou1993}.
The afterglow emission follows after, and is characterized by a power-law decay in multi-wavelength electromagnetic waves \citep{Sari1998ApJ...497L..17S}.
Typically, afterglows are detected 1 hour to 10 days after the prompt burst. 
Traditionally, if a prompt emission lasts longer (shorter) than 2 seconds, the GRB is classified as a long (short) GRB \citep{Kouveliotou1993}.
Primary candidates of the progenitors are collapsars for long GRBs \citep[LGRBs;][]{MacFadyen1999} and binary neutron star mergers for short GRBs \citep[SGRBs;][]{Paczynski1986,Goodman1986ApJ...308L..47G,Eichler1989}
\footnote{Although some recent LGRBs appear to originate from binary neutron star mergers \citep{Rastinejad2022Natur.612..223R,Levan2024Natur.626..737L}, here we consider them as exceptional events and do not focus on them.}.

The power-law decay component in GRB afterglow is naturally explained by the external shock model \citep{Sari1998ApJ...497L..17S}.
However, GRBs often contain excess in their light curves that cannot be explained by this model.
The excess in the form of X-ray flares (XFs) typically begins $\sim 1000$ s after the prompt emission and lasts $\sim 500$ s \citep{Piro2005ApJ...623..314P,Burrows2005Sci...309.1833B,Nousek2006ApJ...642..389N}.
Nearly $\sim30$ \% of long GRBs (LGRBs) have XFs \citep{Yi2016ApJS..224...20Y,Liu2019ApJ...884...59L}.
The light curve of the XFs has a sharp rise and decay, which is inconsistent with the external shock scenario.
This indicates that XFs could originate from the internal dissipation of jets driven by a late central engine activity \citep{Fan2005MNRAS.364L..42F,Ioka2005ApJ...631..429I,Falcone2006ApJ...641.1010F,Liang2006ApJ...646..351L,Zhang2006ApJ...642..354Z}.

Approximately $\sim50$\% of short GRBs (SGRBs) also have X-ray excess, the so-called extended emissions (EEs) and plateau emissions (PEs) \citep{Kisaka2017ApJ...846..142K}.
EE and PE last $\sim300$ s and $\sim10^4$ s after the prompt emission, respectively \citep{Norris2006ApJ...643..266N,Sakamoto2011ApJS..195....2S,Lien2016ApJ...829....7L,Kagawa2015ApJ...811....4K, Kagawa2019ApJ...877..147K,Kaneko2015MNRAS.452..824K}.
They exhibit a flat component followed by a rapid decay in their lightcurves \citep[e.g., see Figure 1 in][]{Kisaka2017ApJ...846..142K}.
These components are also discussed as due to internal dissipations of jets launched by the late activities \citep{Ioka2005ApJ...631..429I,Perna2006ApJ...636L..29P,Zhang2006ApJ...642..354Z,Gao2006ChJAA...6..513G,Metzger2008MNRAS.385.1455M,Rowlinson2013MNRAS.430.1061R,Gompertz2014MNRAS.438..240G,Kisaka2015ApJ...804L..16K}.
However, the physical origin of the late activities and the emission mechanism of the late-time jet are not well understood.

For emissions from internal dissipations of jets, an important characteristic is the distance of the jet's dissipation region from the central engine, the so-called dissipation radius \citep{Rees1994ApJ...430L..93R}.
From small to large dissipation radius, candidate emission scenarios are the dissipative photosphere model \citep{Rees2005ApJ...628..847R}, the canonical internal shock model \citep{Rees1994ApJ...430L..93R}, and the Internal Collision-induced Magnetic Reconnection and Turbulence (ICMART) model \citep{Zhang2011ApJ...726...90Z,Zhang2013PhRvL.110l1101Z}.
These models are often discussed for prompt emissions, and can also be applied to late-time emissions.
However, the dissipation radius cannot be determined by electromagnetic (EM) observations alone, as it is degenerate with the Lorentz factor of the jet.
Constraints on the Lorentz factor for the late-time jet have been estimated in previous work, such as \cite{Troja2015ApJ...803...10T}, \cite{Yi2015ApJ...807...92Y}, and \cite{Matsumoto2020MNRAS.493..783M}.
These works relied on the variability timescale to resolve the degeneracy, but the prescription is only compatible with the internal shock model for the dissipation scenario \citep{Bing2019text}.
Thus, it is important to understand the dissipation radius and the Lorentz factor of the late-time jets as independent parameters.

High-energy neutrino observations could provide
new insight on the parameters of GRB jets.
Neutrinos are emitted by the photohadronic interaction between photons and cosmic rays (accelerated protons) produced by the dissipation of GRB jets 
\citep{Waxman1997PhRvL..78.2292W,Guetta2004APh....20..429G,Murase2006PhRvD..73f3002M,Murase2006PhRvL..97e1101M,Hummer2012PhRvL.108w1101H,Li2012PhRvD..85b7301L,He2012ApJ...752...29H,Kimura2017ApJ...848L...4K,Kimura2022arXiv220206480K,Matsui2023ApJ...950..190M}. 
These works show that the neutrino luminosity depends mainly on the dissipation radius, the Lorentz factor, and the cosmic ray loading factor.
For a given cosmic ray loading factor, it is possible to obtain the dissipation radius and the Lorentz factor with the independent observation of neutrinos and EM waves.
Thus, multi-messenger studies are powerful to probe the physical parameters of the jet.

IceCube, the most sensitive high-energy neutrino observatory at the time of writing, has not detected any neutrino from GRBs yet. 
Still, the non-detection by stacking analyses of $\sim1000$ GRBs sets meaningful constraints on the GRB parameters \citep[e.g.][]{Abbasi2010ApJ...710..346A,Abbasi2011PhRvL.106n1101A,ICCollaboration2012Natur.484..351I,Aartsen2015ApJ...805L...5A,Aartsen2016ApJ...824..115A,Aartsen2017ApJ...843..112A}. 
The non-detection from the brightest GRB, GRB 221009A \citep{Lesage2022BOATGCN,Veres2022BOATGCN,Burns2023BOAT}, put a comparable constraint to the stacking analyses without systematic uncertainty \citep{Abbasi2023BOAT,Murase2022ApJ...941L..10M,Ai2023ApJ...944..115A}.
Also, the extended time window analysis provides constraints on late-time emissions, although they are not as stringent \citep{Abbasi2022grblate}.
Therefore, future cosmic high-energy neutrino detectors are needed, such as IceCube-Gen2 \citep{Aartsen2021JPhG...48f0501A}, KM3Net/ARCA \citep{Aiello2019APh...111..100A}, baikal-GVD \citep{Avrorin2014NIMPA.742...82A}, P-One \citep{Agostini2020NatAs...4..913A}, and TRIDENT \citep{Ye2022arXiv220704519Y} to detect neutrinos from GRBs or put further constraint on GRB-jet parameters.

Neutrino emission during the late activity can be relatively weak due to its weak jet power, but it can be enhanced by an external seed photon field produced by materials around the jet.
A shocked medium, the so-called cocoon, is expected to be formed around jet region by interactions between ambient materials and prompt jets  \citep{Mezaros2001ApJ...556L..37M,Matzner2003MNRAS.345..575M,Toma2007ApJ...659.1420T,Bromberg2011ApJ...740..100B,Nakar2017ApJ...834...28N,Gottlieb2021hydro,Hamidani2020GW2017,Gottlieb2020MHD,Hamidani2021Jetprop,Hamidani2023escapeCoc}.
Recent observations and analysis have reported the direct and indirect signature of the cocoon \citep{Izzo2019Natur.565..324I,Mei2022Natur.612..236M}. 
Therefore, photons or materials of the cocoon can interact with GRB jets, resulting in high-energy emissions \citep{Toma2009ApJ...707.1404T,Shen2010MNRAS.403..229S,Shen2011erratum,Kumar2014MNRAS.445..528K,Kimura2019ApJ...887L..16K}.

\cite{Matsui2023ApJ...950..190M} calculated the neutrino emission from EEs and PEs, taking into account photons from the cocoon, and showed that the enhanced neutrino emission from EEs will be detectable in the next 10 years with IceCube-Gen2.
In addition, \cite{Matsui2023ApJ...950..190M} showed that the detectability is weakly dependent on the Lorentz factor, which is useful to resolve the degeneracy of the dissipation radius and the Lorentz factor.
However, \cite{Matsui2023ApJ...950..190M} calculated the neutrino fluence only for the limited parameters of EE and PE of nearby SGRBs that are likely to be associated with gravitational waves.

In this work, we calculate the detectability of neutrinos from XFs (in LGRBs) and EEs (in SGRBs) for a wide range of parameters, including cosmological GRBs, while taking into account the external seed photon field from the cocoon produced by the prompt jet.
We do not consider PEs because \cite{Matsui2023ApJ...950..190M} shows that the probability of detection is significantly lower.

This paper is organized as follows.
\Sec \ref{sec:jet} describes the formulation to calculate the neutrino spectrum.
\Sec \ref{sec:detect} shows the prospects of neutrino detections by IceCube or IceCube-Gen2 by stacking analysis.
In \Sec \ref{sec:Dis} we discuss constraints on the physical parameters of the late-time jet based on the detectability.
The conclusion of this paper is given in \Sec \ref{sec:Conc}.
Throughout the paper, we use the notation $Q_X = Q/10^X$ in cgs unit unless otherwise noted, and write $Q\prime$ for the physical quantities in the comoving frame of the jet.
We adopt $H_0 = 70\ \rm km \ s^{-1}\ Mpc^{-1}$, $\Omega_{\rm M} = 0.30$, and $\Omega_{\rm \Lambda} = 0.70$ as cosmological parameters, and $z$ represents the redshift.

\section{Photon and Neutrino emission from the late-time jet}\label{sec:jet}
Protons in GRB jets can be accelerated via a dissipation process, where kinetic/magnetic energy is transferred to particles, such as the shock \citep{Rees1994ApJ...430L..93R} or the magnetic reconection \citep{Zhang2013PhRvL.110l1101Z}.
These protons can then interact with surrounding photons, leading to neutrino production \citep{Waxman1997PhRvL..78.2292W}. 
This photohadronic process is the main channel of high-energy neutrino production for the GRBs. 
Here we consider such process.
We take into consideration jets characterized with a dissipation radius $r_{\mathrm{diss}}$, a bulk Lorentz factor $\Gamma$, an isotropic-equivalent luminosity in the X-ray band $L_{\rm X, iso}$, for a duration $t_{\rm dur}$ and with a delay time after the prompt jet $t_{\rm delay}$.
$t_{\rm dur}$ and $t_{\rm delay}$ are defined in the engine rest frame.
The jet model and the dissipation process are described in the following subsections. 

\subsection{photon field}\label{subsec:photon}
Here we estimate the number density of photons in the dissipation region, in order to determine the interaction rate of the photohadronic process. 
We consider two photon components: i) internal photons, and ii) cocoon photons.
\subsubsection{internal photons}
Internal photons are 
produced within the dissipation region of the jet.
Assuming that their energy distribution follows Band function \citep{Band1993ApJ...413..281B}, the (differential) number density of these photons at the dissipation region is estimated as
\begin{eqnarray}
\begin{split}
& \frac{d n^{\prime \mathrm{in}}}{{d \varepsilon^{\prime}_\gamma}} = \\
 & n_{\varepsilon^{\prime}_\gamma,\mathrm{nor}}  \times \begin{dcases}
  \varepsilon^{\prime \alpha}_\gamma \ \mathrm{exp}\left(-\frac{(2+\alpha) \varepsilon^\prime_\gamma}{\varepsilon^{\prime}_{\gamma,\mathrm{pk}}}\right)  \ &(\varepsilon^{\prime}_\gamma \leq \chi \varepsilon^{\prime}_{\gamma,\mathrm{pk}})\\
  \varepsilon^{\prime \beta}_\gamma \  (\chi \varepsilon^{\prime}_{\gamma,\mathrm{pk}})^{\alpha-\beta} \mathrm{exp}(\beta-\alpha) \ &(\varepsilon^{\prime}_{\gamma}>\chi \varepsilon^{\prime}_{\gamma,\mathrm{pk}}),
\end{dcases}
\end{split}
\label{internal}
\end{eqnarray}
where $\alpha$ and $\beta$ are the photon indices at the low- and high-energy regimes, respectively.
$\chi=(\alpha-\beta)/(2+\alpha)$ is a constant, $n_{\varepsilon^{\prime}_\gamma,\mathrm{nor}}$ is the normalization factor, $\varepsilon^{\prime}_{\gamma} $ is the photon energy, and $\varepsilon^{\prime}_{\gamma,\mathrm{pk}}$ is the spectral peak energy.
$n_{\varepsilon^{\prime}_\gamma,\mathrm{nor}}$ is determined so that $L_{\mathrm{X,iso}} =  4 \pi\Gamma^2 r_\mathrm{diss}^2 c \int^{\varepsilon_{\mathrm{X,Max}/\Gamma}}_{\varepsilon_{\mathrm{X,min}/\Gamma}} d\varepsilon^{\prime}_{\gamma} \varepsilon^{\prime}_{\gamma}({dn^{\prime}_\gamma}/{d\varepsilon^{\prime}_{\gamma}} )$ is satisfied, where $c$ is the speed of light. 

We define bolometric luminosity of internal photons as $L_{\gamma,\mathrm{iso}} =  4 \pi\Gamma^2 r_\mathrm{diss}^2 c \int^{\varepsilon^{\prime}_{\gamma,\mathrm{Max}}}_{\varepsilon^{\prime}_{\gamma,\mathrm{min}}} d\varepsilon^{\prime}_{\gamma} \varepsilon^{\prime}_{\gamma}({dn^{\prime}_\gamma}/{d\varepsilon^{\prime}_{\gamma}}) $, where $\varepsilon^{\prime}_{\gamma,\mathrm{min}}$ and $\varepsilon^{\prime}_{\gamma,\mathrm{max}}$ are the minimum and maximum photon energies, respectively.
We set $\varepsilon^{\prime}_{\gamma,\mathrm{min}}= 0.1$ eV and $\varepsilon^{\prime}_{\gamma,\mathrm{max}}= 10^6$ eV, as synchrotron self-absorption and pair creation become effective below and above these energies, respectively \citep{Murase2006PhRvL..97e1101M}.
Note that specific values of $\varepsilon^{\prime}_{\gamma,\mathrm{min}}$ and $\varepsilon^{\prime}_{\gamma,\mathrm{max}}$ have minimal impact on the results.

\subsubsection{photons from the cocoon}
Cocoon photons, are external thermal seed photons provided by the jet heated cocoon \citep{Kimura2019ApJ...887L..16K,Matsui2023ApJ...950..190M}.
Here, the cocoon is defined as the expanding materials with mildly relativistic velocity, whose Lorentz factor is at most a few \citep{Nakar2017ApJ...834...28N,Hamidani2023escapeCoc,Hamidani2023emission}.
This indicates that the maximum velocity of the cocoon is $\sim c$, but its photon field is approximately isotropic in the central-engine rest frame. 

We adopt a one-dimensional (1D) homologously expanding cocoon model, which is updated from the one-zone cocoon model in previous work \citep{Kimura2019ApJ...887L..16K,Matsui2023ApJ...950..190M}.
The number of photons from the cocoon is determined mainly by the temperature.
For a 1D model of the cocoon temperature $T_{\rm coc} (r,t)$, we use 
the temperature distribution shown by 2D relativistic hydrodynamic simulations \citep{Hamidani2023emission}:
\begin{equation}
\label{eq:Tcoc}
\begin{split}
 & a_{\rm rad}T_{\rm coc}^4 (r,t)  = \\ &\frac{3E_{\rm coc, i, br}}{2 \pi (1+3{\rm ln}(c/v_{\rm min}))}
 \times \begin{dcases}
 \frac{r_*}{(v_{\rm min} t)^3ct} & (r \leq v_{\rm min}t ) \\
 \frac{r_*}{r^3 ct} & ( v_{\rm min}t< r < ct ) \\
 0  & (r > ct ),
\end{dcases}
\end{split}
\end{equation}
where $a_{\rm rad}$ is the radiation constant, $v_{\rm min}$ is minimum velocity of the escaped cocoon, $E_{\rm coc, i, br}$ is the total internal energy of the cocoon at the break-out time of the prompt jet, $r_*$ is the radius of the ambient material, $r$ is the distance from the central engine, and $t$ is the time since the break out of the prompt jet.
Spatially integrated internal energy $ \int^{vt}_{0}  2 \pi r^2 dr a_{\rm rad} T_{\rm coc}^4 \propto t^{-1}$ for any $v$ describes the adiabatic expansion.
$E_{\rm coc,i, br}$ is estimated to be 
$E_{\rm coc,i, br} \approx (L_{\rm k,pro,iso}\theta_j^2/4)t_b \eta^\prime$, where $L_{\rm k,pro,iso}$ is the isotropic-equivalent kinetic luminosity of the prompt jet and $\eta^\prime $ is a numerical factor \citep{Hamidani2021Jetprop}.
In order to estimate $L_{\rm k,pro,iso}$, we give the energy fluence ratio between prompt emission and late-time emission as a free parameter: $S_{\rm pro}/ S_{\rm late} = (L_{\rm k, pro,iso}t_{\rm pro})/(L_{\rm k,iso}t_{\rm dur})$, where $t_{\rm pro}$ is the duration of the prompt emission.

We assume that the cocoon photons diffuse into the jet mainly from the material closest to the dissipation region.
The number density of the photons inside the cocoon is usually estimated by the Planck distribution for the temperature $T_{\rm coc}(r_{\rm diss},t_{\rm delay})$.
However, we need to modify it to estimate the number density of the cocoon photons in the jets because of three mechanisms that suppresses the photon number density, as explained below.

The first suppression of the photon number density is by Thomson scattering in the jet.
Here $\tau_j$ is defined as the lateral optical depth of Thomson scattering in the jet.
The number of cocoon photons decay exponentially with respect to $\tau_j$, which can be written as a suppression factor $f_{\tau_j} = e^{-\tau_j}$.
$\tau_j$ is obtained as $\tau_j = (L_{\rm k,iso} \sigma_T \theta_j)/(4\pi r_\mathrm{diss}\Gamma^2m_p c^3)$, where $L_{\rm k,iso} = (L_{\gamma,\rm iso}/f_{\gamma})$, $f_{\gamma}$, $\sigma_T$ and $m_p$ are isotropic-equivalent kinetic luminosity of the jet, the radiation efficiency, the Thomson cross section, and the proton mass, respectively.

The second suppression is by the escape of photons from the cocoon.
Cocoon photons can escape radially for $\tau_{\rm coc}(r,t) <c/v = ct/ r$, where $\tau_{\rm coc}(r,t)$ and $v = r/t$ are the radial optical depth and the expanding velocity of the cocoon at $r$ and $t$.
This leads to its suppression factor $f_{\tau_{\rm coc},\rm esc} = \Theta(\tau_{\rm coc}(r_{\rm diss},t_{\rm delay})-ct_{\rm delay}/r_{\rm diss}) $, where $\Theta(x)$ is the Heaviside step function.

Last suppression is by diffusion process at the surface of the cocoon.
Photons can diffuse into the jet only from the ``skin" of the cocoon adjacent to the jet boundary with a width $s$. 
$s$ can be estimated as the diffusion length within the dynamical time, giving $s \sim (l_{\rm mfp} r_{\rm diss})^{1/2}$, where $l_{\rm mfp}$ is the mean free path for photons in the cocoon. 
Since the photon energy is conserved, the photon energy density is rarefied by the volume ratio between the ``skin" and the jet + ``skin". 
Thus, its suppression factor can be $f_{\tau_{\rm coc},\rm skin} = 1-(r_{\rm diss} \theta_j)^2/(r_{\rm diss} \theta_j + s)^2$.
Eventually, we obtain $f_{\tau_{\rm coc},\rm skin} = [1-(1+\theta_j^{-1} \tau_{\rm coc}^{-1/2} )^{-2}]$ with $r_{\rm diss}/ l_{\rm mfp} = \tau_{\rm coc}$.

In order to estimate $\tau_{\rm coc}(r,t)$ we adopt the 1D density distribution of the cocoon obtaiend by 2D relativistic hydrodynamic simulations \cite{Hamidani2023emission}:
\begin{equation}
\label{eq:rho_coc}
\begin{split}
& \rho_{\rm coc} (r,t)  =   \\ 
& \frac{(p_{\rm coc}-3) M_{\rm coc}}{2\pi(p_{\rm coc}-2) (v_{\rm min}t)^3} 
\times \begin{dcases}
\left(\frac{r}{v_{\rm min}t}\right)^{-2} & (r \leq v_{\rm min}t ) \\
\left(\frac{r}{v_{\rm min}t}\right)^{-p_{\rm coc}} & ( v_{\rm min}t< r < ct )\\
0 &( r > ct ),
\end{dcases}
\end{split}
\end{equation}
where $p_{\rm coc}$ is the density slope and $M_{\rm coc}$ is total mass of the cocoon.
This solution also exhibits the homologous expansion of the cocoon.
$\tau_{\rm coc}(r,t)  = \int^\infty_{r} dr^\prime \rho_{\rm coc}(r^\prime,t) \kappa_{\rm coc} $ leads to
\begin{equation}
\label{eq:taucoc}
\begin{split}
\tau_{\rm coc}(r,t)   = \ & \frac{(p_{\rm coc}-3)\kappa_{\rm coc} M_{\rm coc}}{2\pi(p_{\rm coc}-2) (v_{\rm min}t)^2}  \\
&\times \begin{dcases}
\left(\frac{r}{v_{\rm min}t}\right)^{-1}-\frac{p_{\rm coc}-2}{p_{\rm coc}-1} & (r \leq v_{\rm min}t ) \\
\frac{1}{p_{\rm coc}-1}\left(\frac{r}{v_{\rm min}t}\right)^{1-p_{\rm coc}} & ( v_{\rm min}t< r < ct )\\
0 &( r > ct ),
\end{dcases}
\end{split}
\end{equation}
where $\kappa_{\rm coc}$ is the opacity of the cocoon.

Taking into account $f_{\tau_{\rm j}}$, $f_{\tau_{\rm coc},\rm esc}$, and $f_{\tau_{\rm coc},\rm skin}$, the number density of the cocoon photons in the dissipation region is given by
\begin{equation}
\label{eq:coc}
\begin{split}
&\varepsilon^{\prime}_\gamma \frac{d n^{\prime \mathrm{coc}}}{{d \varepsilon^{\prime}_\gamma}}  = \\ & \frac{8 \pi \varepsilon^{\prime\ 3}_\gamma /(\Gamma^2 h^3 c^3)}{\mathrm{exp}  [\varepsilon^{\prime}_\gamma/(\Gamma k_B T_{\rm coc}(r_{\rm diss},t_{\rm delay}))]-1} \times f_{\tau_j}\ f_{\tau_{\rm coc},\rm esc} f_{\tau_{\rm coc},\rm skin},
\end{split}
\end{equation}
where $h$ and $k_B$ are the Planck constant and the Boltzmann constant, respectively.

The cocoon can be heated by radioactive nuclei in both XFs and EEs.
In LGRBs, ${\rm Ni}^{56}$ nuclei could be synthesized in the cocoon \citep{Tominaga2007ApJ...657L..77T,Barnes2018ApJ...860...38B}. 
In SGRBs, the cocoon is filled with neutron-rich nuclei that do go through beta decay.
In this study, we neglect the heating by radioactive nuclei inside the cocoons for both XFs and EEs, because the temperature and mass density inside the cocoons are too high and low, respectively (e.g.,  see Figure 4 in \citealt{Hamidani2023emission} for EEs). 

\subsection{timescales and cooling rate}\label{subsec:spectrum}
Here we estimate the cooling timescale ($t$) and cooling rate (defined as the inverse of the cooling timescale, $t^{-1}$) for protons in the dissipation region of the late-time jet.

We consider two types of photohadoronic interactions.
One is the photopion process (hereafter as $p\gamma$ process), described as 
\begin{equation}
    p+ \gamma \rightarrow \pi^\pm\ (\pi^0) + X\rightarrow 3\nu+X^\prime,
\end{equation}
where $X$ and $X^\prime$ represent the other particles produced in the $p\gamma$ process in each step.
The other photohadoronic interactions is the Bethe-Heitler process (hereafter as the B-H process), described as 
\begin{equation}
    p+ \gamma \rightarrow e^+ + e^- + p.
\end{equation}
The cooling rates for the $p\gamma$ and the B-H processes are estimated to be
\begin{equation}
t^{\prime-1}_{p\gamma/\mathrm{B-H}} = \frac{c}{2\gamma^{\prime2}_{p}}  \int^{\infty}_{\bar{\varepsilon}_\mathrm{th}} d\bar{\varepsilon}_\gamma \sigma(\bar{\varepsilon}_\gamma) \kappa(\bar{\varepsilon}_\gamma) \bar{\varepsilon}_\gamma \int^{\infty}_{\bar{\varepsilon}_\gamma/{2\gamma_p}} d\varepsilon^{\prime}_\gamma \varepsilon_\gamma^{{\prime}-2}\frac{dn^{\prime}_\gamma}{d\varepsilon^{\prime}_\gamma},
\label{eq;t_pg}
\end{equation}
where $\bar{\varepsilon}_\mathrm{th}$, $\sigma(\bar{\varepsilon}_\gamma)$, $\kappa(\bar{\varepsilon}_\gamma)$, $\gamma_p^\prime = \varepsilon^\prime_\mathrm{p}/(m_\mathrm{p} c^2)$, $\varepsilon_{\gamma}^\prime$, and $dn^\prime_\gamma/d\varepsilon^\prime_\gamma$ are the threshold energy, the cross-section, inelasticity for each reaction in the proton rest frame, the Lorentz factor of protons, the photon energy, and differential number density of photons, respectively. 
For the cross-section and inelasticity of the $p\gamma$ process, we use the fitting formulae based on GEANT4 \citep{Murase2006PhRvD..73f3002M}.
For the B-H process, we use the analytical fitting formulas given in \cite{Stepney1983MNRAS.204.1269S, Chodorowski1992ApJ...400..181C}.
We define $t_{p\gamma, \rm int}$, $t_{\rm B-H,int}$, $t_{p\gamma,\rm coc}$, and $t_{\rm B-H,coc}$ as the cooling timescales using the internal photons and the cocoon photons for two processes, respectively.
We do not need to consider anisotropic effects because protons have an isotropic distribution in the comoving frame of a jet. Due to symmetry, \Eq \ref{eq;t_pg} is valid if either or both photons and protons are isotropic \citep{Murase2014PhRvD..90b3007M}.

We also consider adiabatic cooling, whose timescale is estimated by $t^{\prime}_\mathrm{ad} = r_\mathrm{diss}/(\Gamma c)$.
Synchrotron cooling and proton-proton collision are other cooling process for protons, but we ignore them because they are subdominant for all cases in this work. 
Also, we ignore diffusive escape because its timescale
is much longer than other cooling timescales.

\subsection{Proton and Neutrino Spectrum} \label{subsec:spec}
The observed power-low spectrum of GRB emissions suggests that charged particles are non-thermally accelerated in the dissipation region, although the details of the acceleration process remain uncertain. 
Hence, we consider the proton spectrum (differential number distribution) as
\begin{equation}
\label{eq:dndep}
\frac{dN_p}{d\varepsilon_{p}} = N_{\varepsilon_p,\mathrm{nor}} \left(\frac{\varepsilon_p}{\varepsilon_{p,\mathrm{cut}}}\right)^{-p_\mathrm{inj}} \exp\left(-\frac{\varepsilon_p}{\varepsilon_{p,\mathrm{cut}}}\right),
\end{equation}
where $p_\mathrm{inj}$, $\varepsilon_{p,\mathrm{cut}}$, and $N_{\varepsilon_p,\mathrm{nor}}$ are the power-law index, the proton cutoff energy, and the normalization factor, respectively.  
$N_{\varepsilon_p,\mathrm{nor}}$ is determined by $\xi_p L_{\gamma,\mathrm{iso}} t_\mathrm{dur} =  \int^{\infty}_{\varepsilon_{p,\mathrm{min}}} d\varepsilon_{p} \varepsilon_{p}(dN_p/d\varepsilon_{p})$, where $\xi_p$ is the cosmic ray loading factor \citep{Murase2006PhRvD..73f3002M}.
We use $\varepsilon_{p,\mathrm{min}}=\Gamma\varepsilon^\prime_\mathrm{min}= 3 \Gamma m_p c^2$ as the minimum energy of cosmic ray protons.
The cutoff energy is defined as the energy whose acceleration and cooling timescales are balanced, $t^{\prime-1}_\mathrm{acc}(\varepsilon_{p,\mathrm{cut}}) = t^{\prime-1}_\mathrm{cool}(\varepsilon_{p,\mathrm{cut}})$, where $t^{\prime-1}_\mathrm{cool} = t_{p\gamma}^{\prime -1}+t_{\mathrm{B-H}}^{\prime -1}+t_{\mathrm{ad}}^{\prime -1}$ is the total cooling rate.
The acceleration timescale is obtained by $t^\prime_\mathrm{acc}=\varepsilon^\prime_{p}/(ceB^\prime)$, where $e$ and $B^\prime$ are the elementary charge and the magnetic field strength of the dissipation region, respectively.
$B^\prime$ is estimated by $ B^\prime = \sqrt{(2L_{\gamma,\mathrm{iso}}\xi_B)/(c \Gamma^2 r_\mathrm{diss}^2)}$, where $\xi_B$ is a phenomenological parameter.

We adopt the formulation of neutrino spectra as \citep{Matsui2023ApJ...950..190M}
\begin{equation}
\frac{dN_{\nu_\mu}}{d\varepsilon_{\nu_\mu}} \approx \frac{25}{2}\int d\varepsilon_{\pi} g(\varepsilon_{\pi},\varepsilon_{\nu_\mu})f_{\mathrm{sup},\pi} \left.\left(f_{p\gamma} \frac{dN_p}{d\varepsilon_{p}}\right)\right|_{\varepsilon_{p} = 5\varepsilon_{\pi}},
\label{eq:phinumu}
\end{equation}
for muon neutrinos, and 
\begin{equation}
\begin{split}
\frac{dN_{\bar{\nu}_\mu}}{d\varepsilon_{\bar{\nu}_\mu}} 
    &\approx \frac{dN_{\nu_e}}{d\varepsilon_{\nu_e}} \\
    &\approx \frac{50}{3}\int d\varepsilon_{\mu} g(\varepsilon_{\mu},\varepsilon_{\nu_e})f_{{\rm sup},\mu} \left. \left(f_{{\rm sup},\pi}f_{p\gamma} \frac{dN_p}{d\varepsilon_{p}}\right)\right|_{\varepsilon_{p} = 5\varepsilon_{\pi} = \frac{20}{3}\varepsilon_{\mu}},
\end{split}
\label{phinue}
\end{equation}
for anti-muon neutrinos and electron neutrinos, where $f_{p\gamma}= t^{\prime}_\mathrm{cool}/t^{\prime}_{p\gamma}$, $f_{{\rm sup},i}$, $\varepsilon_i$, and $g(\varepsilon_{i},\varepsilon_{j})d\varepsilon_{j}$ ($i, j$ = $\pi$ or $\mu$) are the pion production efficiency by the $p\gamma$ process, 
the suppression factor by the pion cooling and muon cooling,
energy of the $i$-th particle, and the secondary distribution of $j$-th particle produced by the decay of the parent $i$-th particle of energy $\varepsilon_{i}$, respectively.
As in \cite{Matsui2023ApJ...950..190M}, we approximate $g(\varepsilon_{\pi},\varepsilon_{\nu_\mu}) \approx 4\Theta({\varepsilon_{\pi}}-4\varepsilon_{\nu_\mu})/\varepsilon_{\pi}$ and $g(\varepsilon_{\mu},\varepsilon_{\nu_e}) \approx 3\Theta(\varepsilon_{\mu}-3\varepsilon_{\nu_e})/\varepsilon_{\mu}$. 

$f_{{\rm sup},i}$ stands for the synchrotron loss of pions or muons. 
It is calculated as $f_{{\rm sup},i} = 1- \mathrm{exp}(-t^{\prime}_{i,\mathrm{cool}}/t^{\prime}_{i,\rm dec})$, where $t^{\prime}_{i,\rm dec}$ and $t^{\prime}_{i,\mathrm{cool}}$ represent the lifetime and the cooling timescale of each particle in the comoving frame of the jet, respectively.
The lifetime is expressed as $t^{\prime}_{i,\rm dec} = t_{i,\rm dec} \varepsilon^\prime_i/(m_ic^2)$, where $t_{i,\rm dec}$ and $m_i$ denote the lifetime in the rest frame and the mass of a particle, respectively.
$t^{\prime}_{i,\mathrm{cool}}$ is determined by $t^{\prime-1}_{i,\mathrm{cool}} = t^{\prime-1}_{i,\mathrm{syn}} +t^{\prime-1}_\mathrm{ad} $, with $t^{\prime}_{i,\mathrm{syn}} = 6 \pi m_{i}^4c^3/(m_\mathrm{e}^2\sigma_T B^{\prime 2} \varepsilon^\prime_{ i})$.


The muon neutrino fluences measured on Earth after mixing during the propagation are approximately given by \citep[e.g.,][]{Becker2008}
\begin{equation}
\phi_{\nu_\mu+\bar{\nu}_\mu}  \approx \frac{4}{18}  \phi_{\nu_e+\bar{\nu}_e}^0 +\frac{7}{18}  (\phi_{\nu_\mu+\bar{\nu}_\mu} ^0+ \phi_{\nu_\tau+\bar{\nu}_\tau}^0),
\end{equation}
where $\phi_{i}^0 = ({dN_{i}}/{d\varepsilon_{i}})/(4\pi d_L^2)$ ($i$ = $\nu_e$, $\nu_\mu$, or $\bar{\nu}_{\mu}$) is the neutrino fluence measured on Earth without the flavor mixing, and $d_L$ is the luminosity distance. 

\begin{table*}
  \caption{Fiducial parameters used to calculate neutrino production. The duration ($t_{\rm dur}$), the delay time ($t_{\rm delay}$), X-ray luminosity ($L_{\rm X,iso}$), and the fluence ratio ($S_{\rm pro}/S_{\rm late}$) are obtained from \cite{Yi2015ApJ...807...92Y}, \cite{Liu2019ApJ...884...59L} for XFs, and \cite{Kisaka2017ApJ...846..142K} for EEs. The break out time ($t_{\rm b}$), the radius of the ambient materials ($r_{*}$), the minimum velocity ($v_{\rm min}$), the opacity ($\kappa_{\rm coc}$), the mass ($M_{\rm coc}$) and the density slope ($p_{\rm coc}$) are obtained from \cite{Nakar2017ApJ...834...28N}, \cite{Hamidani2021Jetprop}, \cite{Suzuki2022ApJ...925..148S} for XFs, \cite{Hamidani2021Jetprop}, \cite{Hamidani2023latetime}, \cite{2020MNRAS.496.1369T}, and \cite{2023arXiv230405810B} for EEs. The luminosity distance ($d_L$) and the redshift ($z$) are obtained from \cite{Wanderman2010MNRAS.406.1944W} for XFs, and \cite{Wanderman2015MNRAS.448.3026W} for EEs. $t_{\rm pro}$ is obtained from the bimodal distribution of GRBs as in \cite{Kouveliotou1993}. The value jet opening angle ($\theta_j$) was taken from \citep{2023ApJ...959...13R}.}
  \begin{tabular}{lccccccccc}
    \hline \hline 
    Shared  & $\alpha$ & $\beta$ & $p_\mathrm{inj}$& $\theta_j$ & $f_\gamma$ & $\xi_p$ & $\xi_B$ & $\varepsilon_{\mathrm{X,min}}$, $\varepsilon_{\mathrm{X,max}}$\\
    parameters & &  & & (rad) & & & & (keV) \\
    \hline 
     & $-0.5$ & $-2$ & 2 &0.1 & 0.03 & 10 & 0.33 &  0.3 ,  10 (XRT) \\
    \hline 
  \end{tabular}

  \begin{tabular}{lcccccccccc}
    \hline 
    Parameters  & $\Gamma$&  $r_\mathrm{diss}$ & $t_\mathrm{dur}$ & $t_\mathrm{late}$ & $L_\mathrm{X,iso}$  & $\varepsilon_{\gamma,\mathrm{pk}}$ & $d_L$ & $z$ \\
     of the emission  &  &  (cm)& (s)& (s) & (erg/s)   & (keV) & (Gpc)  \\
    \hline 
    XF  &  100  & $10^{13}$& $650$& $1300$ & $1.0\times10^{48}$  & 0.3 & 26 & 3.1   \\
    EE  &  200 & $10^{12}$ & $300$ & $300$ &$1.2\times10^{48}$  & 10 & 5.8 & 0.9 \\
    \hline 
    \hline 
      Parameters &$S_{\rm pro}/S_{\rm late}$ &$t_{\rm pro}$ & $t_b$& $\eta^\prime$ & $r_*$& $v_{\rm min}$ & $\kappa_{\rm coc}$& $M_{\rm coc}$& $p_{\rm coc}$\\
      of the cocoon& & (s) & (s) &  & (cm) & $c$ & (${\rm cm^2}/g)$& ($M_{\odot}$)& \\
      \hline 
      XF  & 100 &10 & 3.6 &0.5 & $4.0\times10^{10}$& 0.01 & 0.03 & 1.0 & 5\\
      EE  &1 &1 & 1.7 & 0.25 & $1.7\times10^9$& 0.3 & 1.0 & $10^{-4}$ & 8\\
      \hline 
    
  \label{fiducial parameters}
  \end{tabular}
\label{tab:param}
\end{table*}

\begin{figure*}\hspace{-1cm}
    \begin{tabular}{cc}
      \begin{minipage}[t]{0.5\hsize}
        \centering
        \includegraphics[keepaspectratio, scale=0.43]{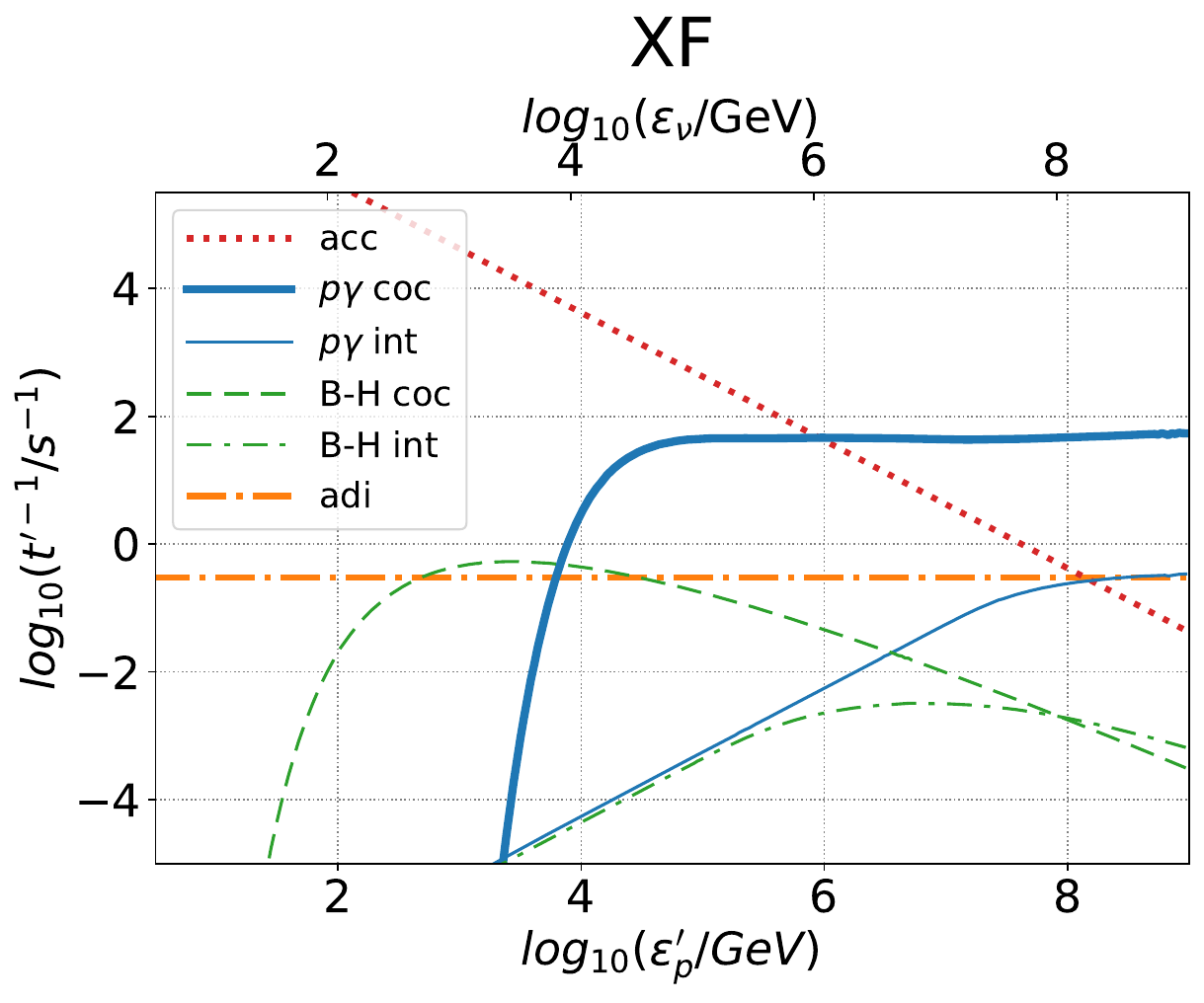}
      \end{minipage}
      
      \begin{minipage}[t]{0.5\hsize}
        \centering
        \includegraphics[keepaspectratio, scale=0.43]{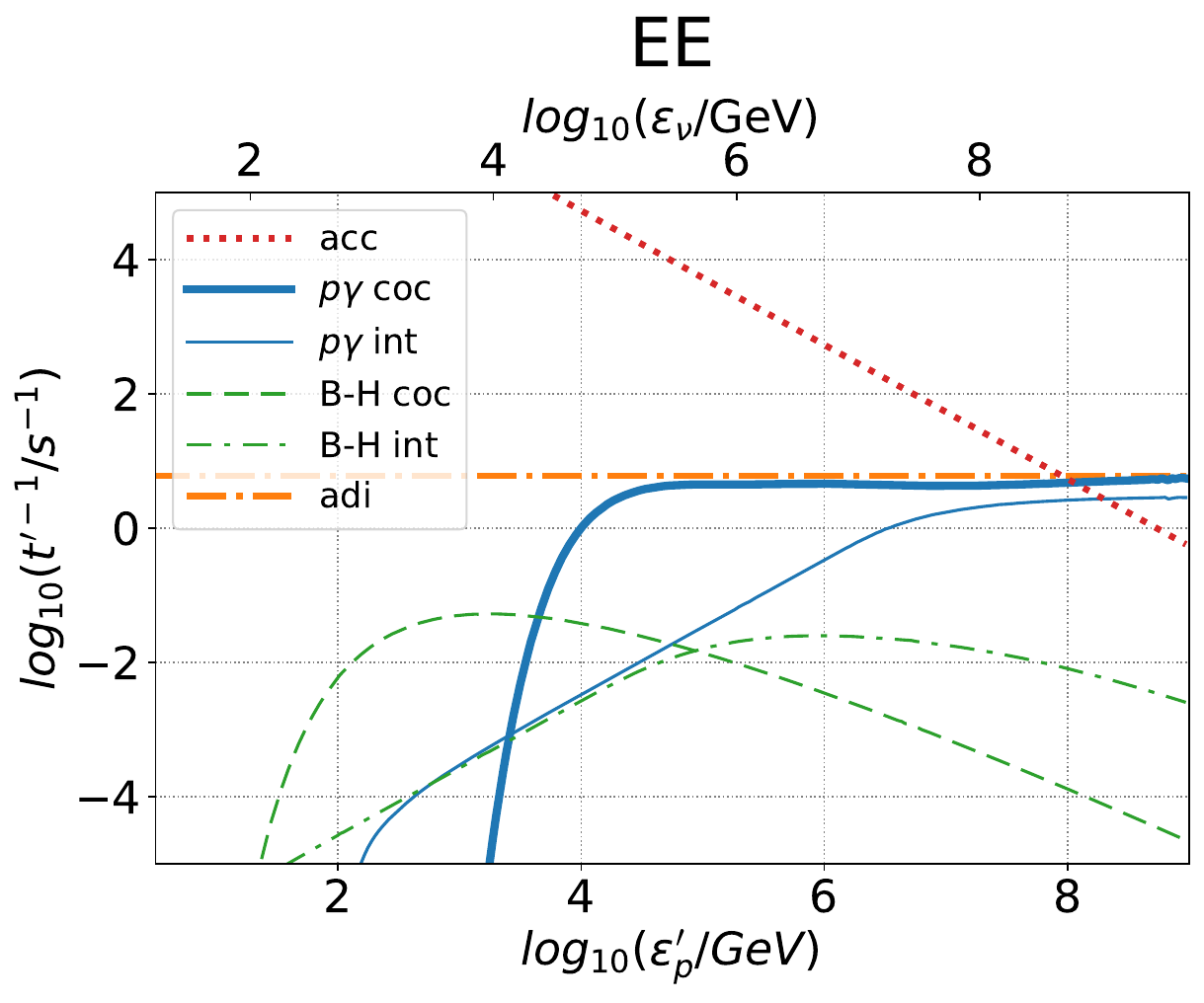}
      \end{minipage}
      
    \end{tabular}
    \caption{The cooling and acceleration rate for our fiducial parameters of XF (in LGRBs; left) and EE (in SGRBs; right). Red thick dotted line is for the acceleration rate, blue thick (thin) solid line is for the cooling rate of the $p\gamma$ process by cocoon (internal) photons, $t^{\prime -1}_{p\gamma, \rm coc}$ ($t^{\prime -1}_{p\gamma, \rm int}$), green thin dashed (dotted-dashed) line is for the cooling rate of the B-H process by cocoon (internal) photons, $t^{\prime -1}_{\rm B-H, coc}$ ($t^{\prime -1}_{ \rm B-H, int}$), and orange thick dotted-dashed line is for the adiabatic cooling rate, $t^{\prime -1}_{\rm ad}$.
}
    \label{fig:time_fid}
  \end{figure*}

\begin{figure*}\hspace{-1cm}
    \begin{tabular}{cc}
      \begin{minipage}[t]{0.5\hsize}
        \centering
        \includegraphics[keepaspectratio, scale=0.4]{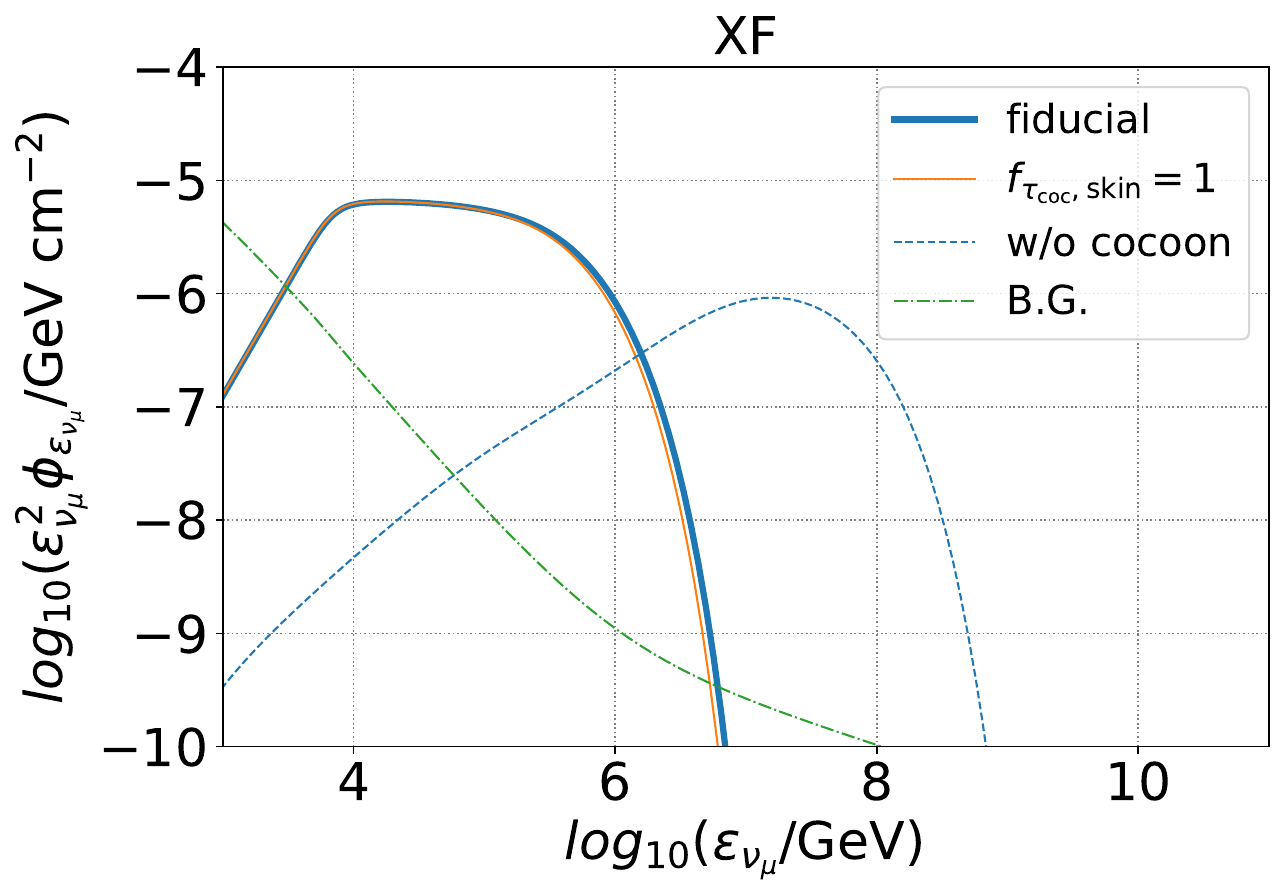}
      \end{minipage}
      
      \begin{minipage}[t]{0.4\hsize}
        \centering
        \includegraphics[keepaspectratio, scale=0.4]{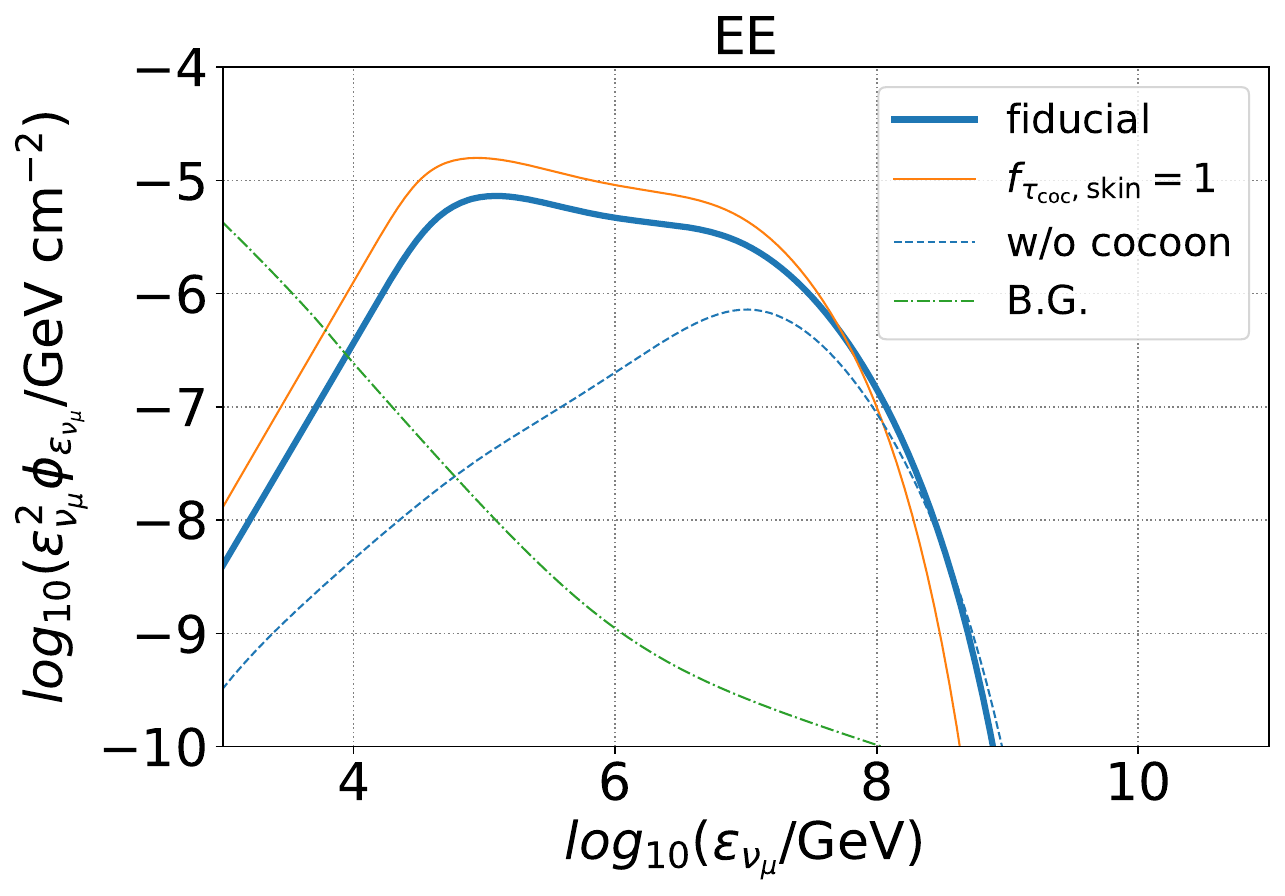}
      \end{minipage}
      
    \end{tabular}
    \caption{The neutrino fluence for 
    XF (left) and EE (right). The blue thick solid is for our fiducial parameters, the blue thin dashed line is the fluence without taking into account cocoon photons, and the orange thin solid line is the fluence for an efficient diffusion case, $f_{\tau_{\rm coc},\rm skin} =1$ discussed in \Ap \ref{app:diffusion}. The green dotted-dashed line is for the all-sky averaged atmospheric and astronomical background within 3 $\rm deg^2$ and 300 s, as described in \Sec \ref{sec:detect}.
}
    \label{fig:fluence_fid}
  \end{figure*}

\Fg \ref{fig:time_fid} and \ref{fig:fluence_fid} show the cooling rate, acceleration rate, and neutrino fluence with our fiducial parameters from \Tb \ref{tab:param}. 
For our fiducial case of XF (left panels, $v_{\rm min}t_{\rm delay}<r_{\rm diss}<ct_{\rm delay}$), $t^{\prime -1}_{p\gamma, \rm coc}$ is the highest cooling rate for protons with $\varepsilon_p^\prime > 100$ TeV, as illustrated in \Fg \ref{fig:time_fid}.
In this case, the cocoon temperature is estimated as $k_B T_{\rm coc} = 22\ E_{\rm coc, i, 51}^{1/4} r_{\rm diss, 13}^{-3/4} r_{*, 10.5}^{1/4} t_{\rm delay, 3}^{-1/4}\ \mathrm{eV} $.
The threshold energy of protons for the $p\gamma$ process with typical cocoon photons is
$\varepsilon^{\prime}_{p, \rm thrsh} = (\bar{\varepsilon}_\mathrm{th} m_pc^2)/(2 k_B T_{\mathrm{coc}} \Gamma) = 42\  E_{\rm coc, i, 51}^{-1/4} r_{\rm diss, 13}^{3/4} r_{*, 10.5}^{-1/4} t_{\rm delay, 3}^{1/4} \Gamma_2^{-1}$ TeV.
This value aligns with the peak energy of $t^{\prime -1}_{p\gamma, \rm coc}$ shown in \Fg \ref{fig:time_fid}.
Substituting $k_B T_{\rm coc}$ into \Eq (A5) in \cite{Matsui2023ApJ...950..190M} yields $t^{\prime -1}_{p\gamma, \rm coc} = 54\ \ E_{\rm coc, i, 51}^{3/4} r_{\rm diss, 13}^{-9/4} r_{*, 10.5}^{3/4} t_{\rm delay, 3}^{-3/4} \Gamma_2\ \mathrm{s^{-1}}$ above the threshold.
This value explains $t^{\prime -1}_{p\gamma, \rm coc}$ for $\varepsilon_p^\prime > 100$ TeV.
The suppression factor due to the shielding of the jet is small, given $\tau_j = 0.32 \  L_{k,\mathrm{iso},50.5} r_{\mathrm{diss},13}^{-1}\Gamma_{2}^{-2}\theta_{j,-1} <1$.
In addition, $\tau_{\rm coc} = 14\ \kappa_{\rm coc, -1.5} M_{\rm coc,0} \beta_{\rm min,-2}^2 r_{\rm diss,13}^{-4} t_{\rm delay, 3}^2 > 1$, and $f_{\tau_{\rm coc},\rm skin} \simeq 0.93$ efficiently supply cocoon photons, where $M_{\rm coc,n}$ = $M_{\rm coc}/(10^{n}M_\odot$).

Also for the fiducial case of EEs (right panels, $r_{\rm diss}<v_{\rm min}t_{\rm delay}$), $t^{\prime -1}_{p\gamma, \rm coc}$ is the highest (comparable to $t^{\prime -1}_{\rm ad}$) for $\varepsilon_p^\prime > 100$ TeV.
The values related to $t^{\prime -1}_{p\gamma, \rm coc}$ are 
$k_B T_{\rm coc} = 25\ E_{\rm coc, i, 50}^{1/4}\ \beta_{\rm min,-0.5}^{-3/4}\ r_{*, 9}^{1/4}\ t_{\rm delay, 2.5}^{-1}\ \mathrm{eV} $, 
$\varepsilon^{\prime}_{p,\rm thrsh} = 18\ E_{\rm coc, i, 50}^{-1/4}\ \beta_{\rm min, -0.5}^{3/4}\ r_{*, 9}^{-1/4}\ t_{\rm delay, 2.5} (\Gamma/200)^{-1}$ TeV, 
$t^{\prime -1}_{p\gamma, \rm coc} = 86 \ E_{\rm coc, i, 50}^{3/4} \beta_{\rm min, -0.5}^{-9/4} r_{*, 9}^{3/4} t_{\rm delay, 2.5}^{-3} (\Gamma/200)\ \mathrm{s^{-1}}$ for $\varepsilon_p^\prime > \varepsilon^{\prime}_{p,\rm thrsh}$ (without suppression), 
$\tau_j = 0.80 \  L_{k,\mathrm{iso},50.5} r_{\mathrm{diss},12}^{-1}(\Gamma/200)^{-2}\theta_{j,-1} \sim1$, $\tau_{\rm coc} = 8.2\times 10^3 \ \kappa_{\rm coc,0} M_{\rm coc,-4} \beta_{\rm min,-0.5}^{-1} r_{\mathrm{diss},12}^{-1} t_{\rm delay, 2.5}^{-1}$, and $f_{\tau_{\rm coc},\rm skin} \simeq 0.19$.
They exhibit a behavior similar to that of XFs, except for a factor of 10 suppression due to $f_{\tau_{\rm j}}\sim0.5$ and $f_{\tau_{\rm coc},\rm skin}\sim0.2$.

Therefore, neutrinos are mainly produced by the $p\gamma$ process with cocoon photons. 
\Fg \ref{fig:fluence_fid} also shows the fluence without cocoon photons (blue dashed lines).
We confirm that the neutrino emission is strongly enhanced by cocoon photons, especially at $\varepsilon_{\nu_\mu} \sim 100$ TeV, where IceCube is the most sensitive. 
The spectrum and its peak for EEs are also consistent with \cite{Matsui2023ApJ...950..190M}. 
The other orange lines in \Fg \ref{fig:fluence_fid} are the lines for $f_{\tau_{\rm coc},\rm skin} =1$ mentioned in \Ap \ref{app:diffusion}.

Moreover, the neutrino luminosity depends weekly on $\Gamma$.
The neutrino luminosity is approximately estimated as $L_\nu \sim L_p f_{p\gamma} \approx L_p t^{\prime-1}_{p\gamma}/t^{\prime-1}_{\rm cool}$ , where $L_p = \xi_p L_{\gamma, \rm iso}$ is the proton luminosity \citep[e.g.][and references therein]{Meszaros2015review,Kimura2022arXiv220206480K}.
For our fiducial parameters, with cocoon photons, $f_{p\gamma} \sim t^{\prime-1}_{p\gamma}/t_{\rm ad}^{\prime -1} \sim 1$ is satisfied for a wide range of proton energy.
$t^{\prime-1}_{p\gamma, \rm coc}$ is proportional to $\Gamma$ as well as $t_{\rm ad}^{\prime -1}\propto \Gamma $, because the number density of external seed photons increases by $\Gamma$ at the jet comoving frame due to the Lorentz contraction.  
This results in $t^{\prime-1}_{p\gamma}/t_{\rm ad}^{\prime -1} \gtrsim 1$ independently of $\Gamma$, leading to $f_{p\gamma} \sim 1$ regardless of the value of $\Gamma$ around the fiducial value. 
Furthermore, the peak energy of neutrinos is obtained by $\sim \varepsilon^{\prime}_{p,\rm thrsh} \Gamma/20 \propto \Gamma^0$.
Thus, the peak value and the peak energy of the neutrino luminosity are independent of $\Gamma$ as long as cocoon photons can contribute to the neutrino production.
This behavior is useful in constraining the parameters of the jet.

On the other hand, if we ignore the cocoon photons, $t_{p\gamma,\rm int} ^{-1}\propto 1/\Gamma$ (not as $t_{p\gamma,\rm coc} ^{-1}\propto t_{\rm ad} ^{-1}\propto \Gamma$) because the number density of the internal photons is deboosted.
Thus, the pion production rate by internal photons, $t_{p\gamma,\rm int}^{-1}/ t_{\rm ad} ^{-1} \propto 1/\Gamma^2$, should be much lower than unity for a high value of $\Gamma$. 
$f_{p\gamma} \sim 1$ for high $\Gamma$ is the result with cocoon photons.

\section{Detectability} \label{sec:detect}
This section presents the detectability of neutrinos calculated in \Sec \ref{sec:jet}.
The expected number of $\nu_\mu$-induced events for a single source is estimated as 
\begin{equation}
\label{eq:exp number}
\bar{N}_{\nu_\mu} = \int d \varepsilon_{\nu_\mu} \phi_{\nu_\mu+\bar{\nu}_\mu} (\varepsilon_{\nu_\mu}) A_\mathrm{eff}(\delta,\varepsilon_{\nu_\mu}), 
\end{equation}
where $A_\mathrm{eff}$ is the effective area for a detector, and $\delta$ ($-90^\circ \leq \delta < 90^\circ$) is the declination angle. 
We employ the 10-year point source analysis of IceCube \citep{ICCollaboration2021arXiv210109836I} for $A_{\rm eff}$ and assume that the effective area of IceCube-Gen2 is 5 times larger than that of IceCube \citep{Aartsen2021JPhG...48f0501A}.
For our fiducial parameters and with IceCube-Gen2, $\langle \bar{N}_{\nu_\mu} \rangle_\delta = \int d \delta {\rm cos} \delta  \bar{N}_{\nu_\mu}/2 = 4.4 \times 10^{-4}$ (for XFs) and $\langle \bar{N}_{\nu_\mu} \rangle_\delta = 4.8 \times 10^{-4}$ (for EEs).
This indicates that we need to stack $\sim 10^3$ of GRBs to detect neutrinos.

The following describes prospects for detectability when stacking $N_{\rm GRB}$ GRBs.
The expected number of neutrino detections for the stacking analysis is obtained by 
\begin{equation}
    \bar{N}_{\nu_\mu,\rm stacked}  = \langle \bar{N}_{\nu_\mu} \rangle N_{\rm GRB},
\end{equation}
where
\begin{equation}
    \langle \bar{N}_{\nu_\mu} \rangle =\frac{\int d\Omega d\chi \chi^2 R(z) \bar{N}_{\nu_\mu}/(1+z)}{\int d\Omega d\chi \chi^2 R(z) /(1+z)},
\end{equation}
is the expected number of neutrinos averaged over the volume, $\chi(z)$ is the comoving distance, and $R(z)$ is the event rate of GRBs.
We use 
\begin{eqnarray}
\begin{split}
R(z) = & 1.3\ {\rm Gpc^{-3}yr^{-1}}\\
& \times \begin{cases}
(1+z)^{2.1} & ( z <  3.1 ) \\
4.1^{2.1}\times (1+z)^{-1.4} & ( z \geqq 3.1 ),
\end{cases}
\end{split}
\end{eqnarray}
for LGRB \citep{Wanderman2010MNRAS.406.1944W}, and
\begin{eqnarray}
\begin{split}
     R(z) = &4.1\ {\rm Gpc^{-3}yr^{-1}} \\ 
     & \times\
      \begin{cases}
        \mathrm{exp}[(z-0.9)/0.39] & ( z <  0.9 ) \\
        \mathrm{exp}[-(z-0.9)/0.26] & ( z \geqq 0.9 ),
      \end{cases}
\end{split}
\end{eqnarray}
for SGRB \citep{Wanderman2015MNRAS.448.3026W}.
These event rates are not beam-corrected.
We use them because neutrinos are beamed to the jet axis as GRB emissions.

\begin{figure*}\hspace{-1.5cm}
    \begin{tabular}{cc}
      \begin{minipage}[t]{0.5\hsize}
        \centering
        \includegraphics[keepaspectratio, scale=0.6]{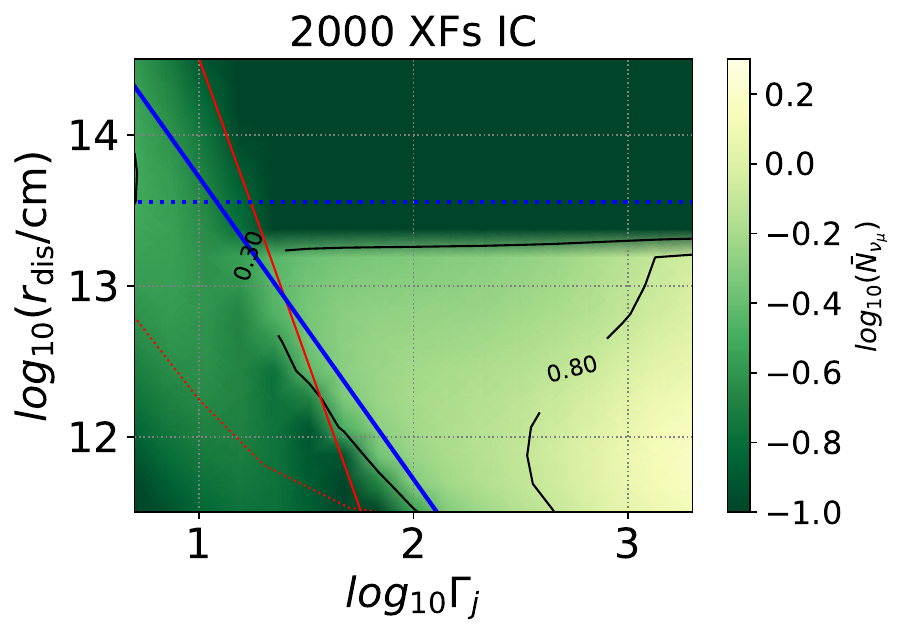}
      \end{minipage} 
      \begin{minipage}[t]{0.45\hsize}
        \centering
        \includegraphics[keepaspectratio, scale=0.6]{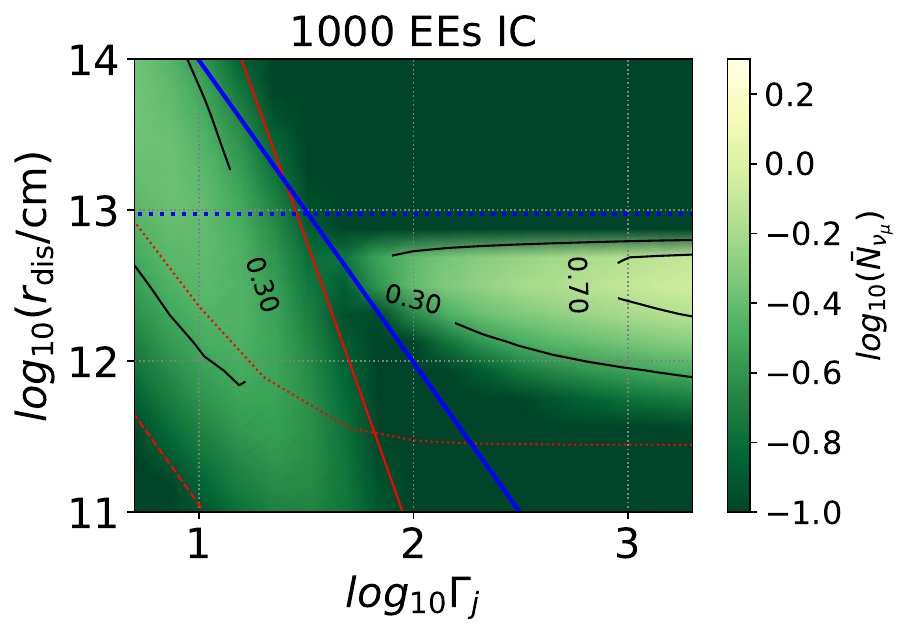}
      \end{minipage}\\
      
      \begin{minipage}[t]{0.5\hsize}
        \centering
        \includegraphics[keepaspectratio, scale=0.6]{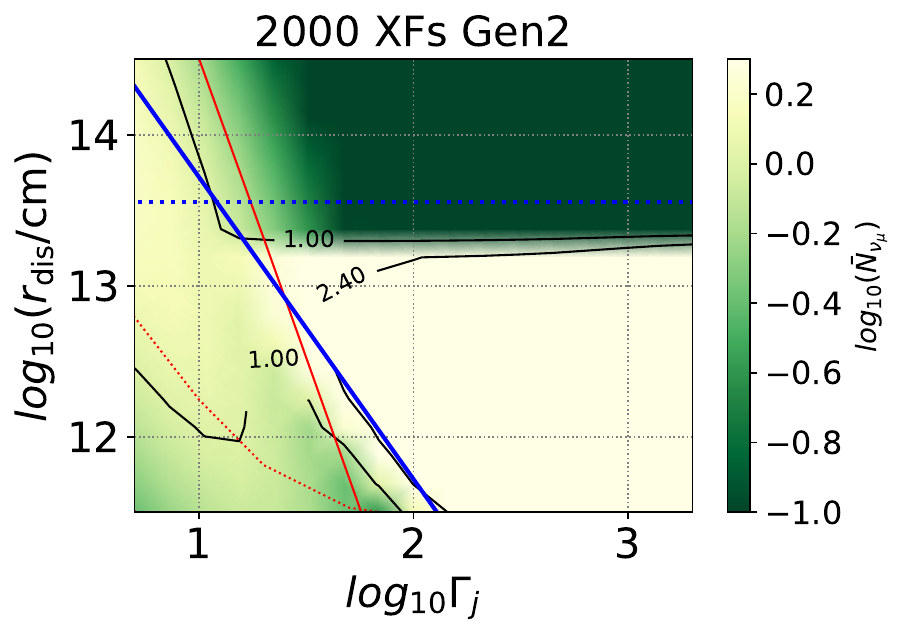}
      \end{minipage}
      
      \begin{minipage}[t]{0.45\hsize}
        \centering
        \includegraphics[keepaspectratio, scale=0.6]{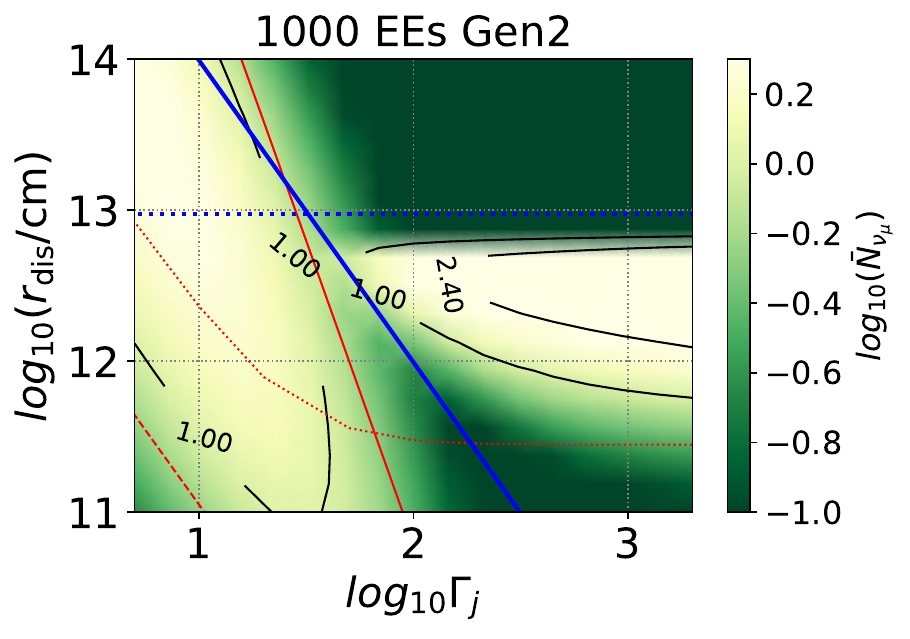}
      \end{minipage}
    \end{tabular}
    \caption{Color maps of the expected number of neutrino detection for 2000 XFs (left) and 1000 EEs (right) with IceCube (top) or IceCube-Gen2 (bottom). Black solid lines are the contours for $\bar{N}_{\nu_\mu,\rm stacked}$. The solid red lines represent the condition given by \Eq \eqref{eq;fast_pi_pro_int}, where the pions are sufficiently produced. The dashed (dotted) red lines represent the condition given by \Eq \eqref{eq;fast_pi_cool} (\Eq \eqref{eq;fast_mu_cool}), where pions (muons) are significantly cooled. For XFs, the red dashed lines are outside of the plots. The solid blue lines represents the condition given by \Eq \eqref{eq;coc_shield}, where cocoon photons can enter into the dissipation region. The dotted blue lines represents the condition given by \Eq \eqref{eq:cocsize}, where the cocoon can cover the dissipation region.
}
    \label{fig:number_fid}
  \end{figure*}

\Fg \ref{fig:number_fid} shows $\bar{N}_{\nu_\mu,\rm stacked}$ for XFs ($N_{\rm GRB} =2000$) and EEs ($N_{\rm GRB} =1000$).
As can be seen in the top panels, the expected numbers for IceCube are less than 2.4. 
This means that we cannot put any constraint on jet physical parameters with 90\%-confidence level in the case of non-detection \citep{Feldman1998PhRvD..57.3873F}.
The bottom panels of \Fg \ref{fig:number_fid} show that the expected number for IceCube-Gen2 is higher than 2.4 in the high $\Gamma$ and relatively low $r_{\rm diss}$ regions.
The contour of $\bar{N}_{\nu_\mu,\rm stacked} = 2.4$ will mimic the 90 \% confidence limit if IceCube-Gen2 does not detect neutrinos \citep{Feldman1998PhRvD..57.3873F}.

The region of high detectability is explained by \Eq (16)-(18) in \cite{Matsui2023ApJ...950..190M}. Under the red solid lines, given by
\begin{equation}
\begin{split}
 r_\mathrm{dis,12} \leq 6.0 &\  \Gamma_{2}^{-2}L_{\gamma,\mathrm{iso},50}  \left(\frac{\varepsilon_{\gamma,\mathrm{pk}}}{100\ \mathrm{keV}}\right)^{-1}\\
    &\times \left[1+5.4\ \left(\frac{\varepsilon_{\gamma, \mathrm{pk}}}{100\ \mathrm{keV}}\right)\ \Gamma_{2}^{2}\right]^{-1},
\end{split}
 \label{eq;fast_pi_pro_int}
\end{equation}
internal photons are expected to sufficiently produce pions that decay into neutrinos with $\sim$ 100 TeV \footnote{As a note, \Eq (16) in \citep{Matsui2023ApJ...950..190M} includes a typo where the factor 0.54 should be 5.4.}.
Additionally, the $p\gamma$ process with internal photons should avoid significant pion cooling suppression for sufficient neutrino production. 
This condition is indicated by the red dashed lines\footnote{For XFs (left panels), this line is outside of the plots.}, expressed as
\begin{equation}
 r_\mathrm{dis,12} \geq 4.6\times10^{-3}\  \Gamma_{2}^{-2}L_{\gamma,\mathrm{iso},50}^{1/2}\left(\frac{\xi_\mathrm{B}}{0.33}\right)^{1/2} +  2.5\times10^{-3}.
 \label{eq;fast_pi_cool}
\end{equation}
Also, muon cooling is significant under the red dotted lines, given by
\begin{equation}
 r_\mathrm{dis,12} \geq 8.5\times10^{-2}\  \Gamma_{2}^{-2}L_{\gamma,\mathrm{iso},50}^{1/2}\left(\frac{\xi_\mathrm{B}}{0.33}\right)^{1/2} +  0.28,
 \label{eq;fast_mu_cool}
\end{equation}
but this effect changes the result at most by a factor of 3.

Even in the region above the red lines of \Eq \eqref{eq;fast_pi_pro_int}, cocoon photons can effectively produce neutrinos when $f_{\tau_{\rm coc},\rm esc } = 1$ and $f_{\tau_j} = e^{-\tau_j} < $ 1/10.
The former conditions are usually met if dissipation region is covered by the cocoon,
\begin{equation}
\label{eq:cocsize}
    r_{\rm diss} < R_{\rm coc} = ct_{\rm delay}, 
\end{equation}
as indicated by the dotted blue lines. 
The latter conditions are above the blue solid lines, expressed as
\begin{equation}\label{eq;coc_shield}
    r_\mathrm{dis,12} \geq  40\times \Gamma_{2}^{-2} L_{\gamma,\mathrm{iso},50}\theta_{j,-1} f_{\gamma,-1.5}^{-1} \times \ln10.
\end{equation}

The suppression factor by the diffusion process ($f_{\tau_{\rm coc},\rm skin}$) does not reduce the expected number of the neutrino detection for XFs. 
When $\tau_{\rm coc} \gg \theta_j^{-2}$, $f_{\tau_{\rm coc},\rm skin}$ decreases as $f_{\tau_{\rm coc},\rm skin}\sim \theta_j^{-1} \tau_{\rm coc}^{-1/2}$.
For XFs, $\tau_{\rm coc} \propto r_{\rm diss}^{-4}$ leads to $f_{\tau_{\rm coc},\rm skin}\propto r_{\rm diss}^{2}$ for $r_{\rm diss} > v_{\rm min}t_{\rm delay}$.
The decrease of $f_{\tau_{\rm coc},\rm skin}$ for a smaller $r_{\rm diss}$ is, however, canceled by the increase of photon number density, $n_\gamma \propto T_{\rm coc}^3\propto r_{\rm diss}^{-9/4}$.
Thus, $t^{\prime-1}_{p\gamma}$ is not significantly affected by $f_{\tau_{\rm coc},\rm skin}$. 

For EEs, $f_{\tau_{\rm coc},\rm skin}$ decrease the expected number of the neutrino detection.
$f_{\tau_{\rm coc},\rm skin}\sim  \tau_{\rm coc}^{-1/2} \propto r_{\rm diss}$ decreases with small $r_{\rm diss}$ while $n_\gamma \propto T_{\rm coc}^3\propto r_{\rm diss}^{0}$ is constant for $r_{\rm diss} < v_{\rm min}t_{\rm delay}$.
This suppresses the detection numbers at $r_{\rm diss} \sim 10^{11}$ cm and $ \Gamma>100$.

We must consider noise levels when constraining GRB parameters based on the (non-)detection of neutrinos.
The expected number of background neutrinos, denoted as $\bar{N}_{\nu_\mu,\rm BG}$, is determined by substituting $\phi_{\nu_\mu+\bar{\nu}_\mu} = \phi_{\rm astr} + \phi_{\rm atomos}$ into \Eq \eqref{eq:exp number}. Here, $\phi_{\rm astr/atomos}$ represents the background neutrino fluence originating from astrophysical or atmospheric sources within the time and spatial window corresponding to GRB late-time emissions.
These values are calculated as $\phi_{\rm astr/atomos} = \Phi_{\rm astr/atomos} t_{\rm dur} \Delta \Omega$, where $\Phi_{\rm astr/atomos}$ and $\Delta \Omega$ denotes the background neutrino flux for each source and a typical uncertainty of a neutrino direction, respectively.

For $\Phi_{\rm astr}$, we use the expression $\Phi_{\rm astr} = 1.44\times 10^{-18} (\varepsilon_{\nu_\mu}/100\ \mathrm{TeV})^{-2.37}\ \rm GeV^{-1}\ cm^{-2}\ str^{-1}\ s^{-1}$, as provided by \cite{Abbasi2022ApJ...928...50A}.
For $\Phi_{\rm atomos}$, we refer to \cite{Honda2007PhRvD..75d3006H}, which estimates $\Phi_{\rm atomos}$ only below $\varepsilon_{\nu_\mu} =$ 10 TeV. 
We extrapolate it by a power-law function above this energy as a conservative choice.
The extrapolation does not significantly impact the results as it is dominated by the astrophysical background in $\varepsilon_{\nu_\mu} \geq$ 100 TeV.

To be precise, $\Delta \Omega$ should be the larger value of either a typical directional uncertainty for GRB or neutrino detections.
Fermi-GBM, the most efficient GRB detector \citep{vonKienlin2020ApJ...893...46V}, has poor angular resolution $\gtrsim 10 \ \rm deg^2$.
However, Einstein Probe \citep{Yuan2015EinsteinProbe} and nanosatellite constellation, such as CAMELOT \citep{Werner2018CAMELOT} and HERMES \citep{Fuschino2019HERMES}, will improve the resolution $\lesssim 1 \ \rm deg^2$ in the near future.
On the other hand, IceCube has an angular resolution $\sim 3\ \rm deg^2$ for 1 TeV.

The expected number of background events for stacked GRBs is obtained by
\begin{equation}
\begin{split}
\bar{N}_{\nu_\mu,\rm st,BG} 
    &= N_{\rm GRB} \langle \bar{N}_{\nu_\mu, \rm BG} \rangle_\delta\ \\
    &=N_{\rm GRB} \int \frac{d\delta {\rm cos} \delta}{2} \bar{N}_{\nu_\mu, \rm BG}(\delta),
\end{split}
\label{eq:BG}
\end{equation}
resulting in $\bar{N}_{\nu_\mu,\rm st,BG} = 0.3 - 0.5$ for IceCube-Gen2 with $t_{\rm dur}=300\ \mathrm{s}$, $\Delta\Omega = 2 - 4\ \mathrm{deg}^2$, and $N_{\rm GRB}=2000$.
Its background contamination probability is approximately 30 \%.
This background changes 90\% CL Feldman-Cousins upper limit to $\bar{N}_{\nu_\mu, \rm stacked} \sim 2.5 - 2.8$ \citep{Feldman1998PhRvD..57.3873F}.

However, the background noise can be cut by setting the energy threshold $\varepsilon_{\nu_\mu, \rm th}$ for neutrino searches.
Choosing $\varepsilon_{\nu_\mu, \rm th}$ = 10 TeV and 100 TeV yields $\bar{N}_{\nu_\mu,\rm st,BG} = (0.7 - 1.3) \times 10^{-2} $ and $\bar{N}_{\nu_\mu,\rm st,BG} = (3.5 - 6.9)\times 10^{-4} $, respectively, with same value of $t_{\rm dur}, \ \Delta \Omega$, and $N_{\rm GRB}$.
With these background, 90\% CL upper limit is still on $\bar{N}_{\nu_\mu, \rm stacked} \simeq 2.4$ \citep{Feldman1998PhRvD..57.3873F}.

\begin{figure*}\hspace{-1.5cm}
    \begin{tabular}{cc}
      \begin{minipage}[t]{0.5\hsize}
        \centering
        \includegraphics[keepaspectratio, scale=0.6]{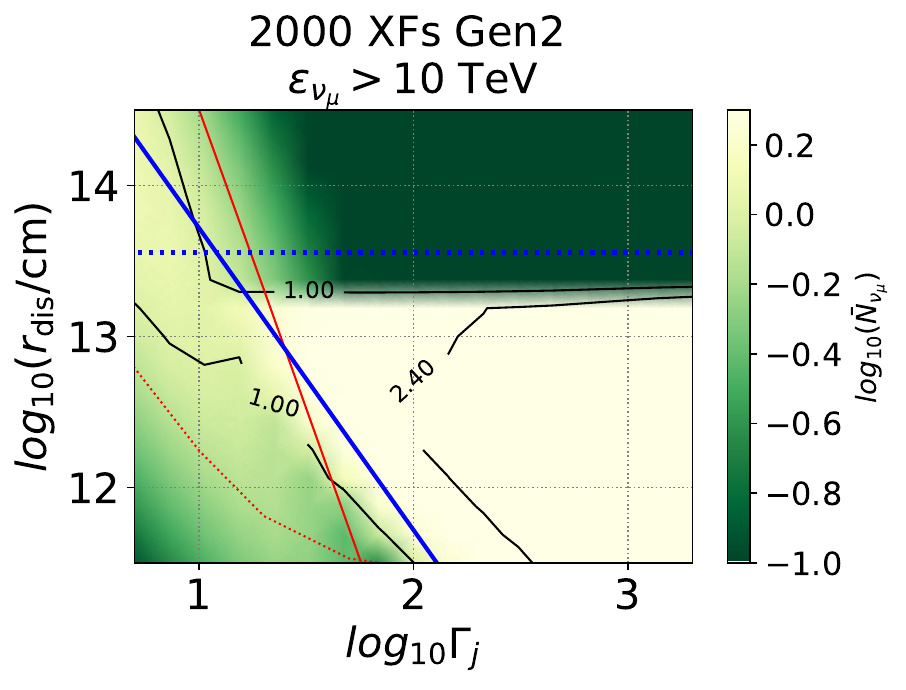}
      \end{minipage} 
      \begin{minipage}[t]{0.45\hsize}
        \centering
        \includegraphics[keepaspectratio, scale=0.6]{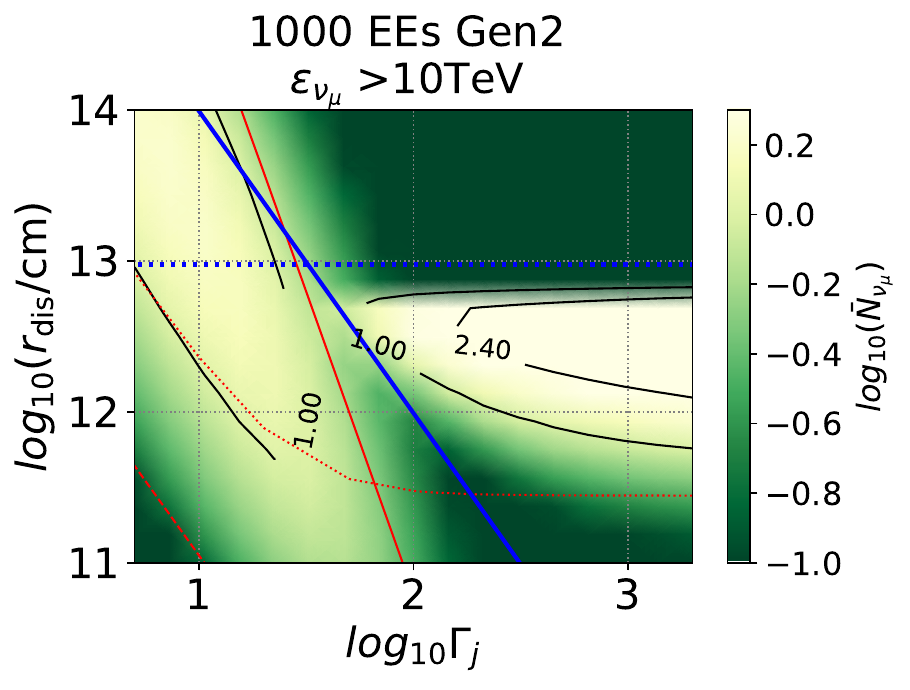}
      \end{minipage}\\
      
      \begin{minipage}[t]{0.5\hsize}
        \centering
        \includegraphics[keepaspectratio, scale=0.6]{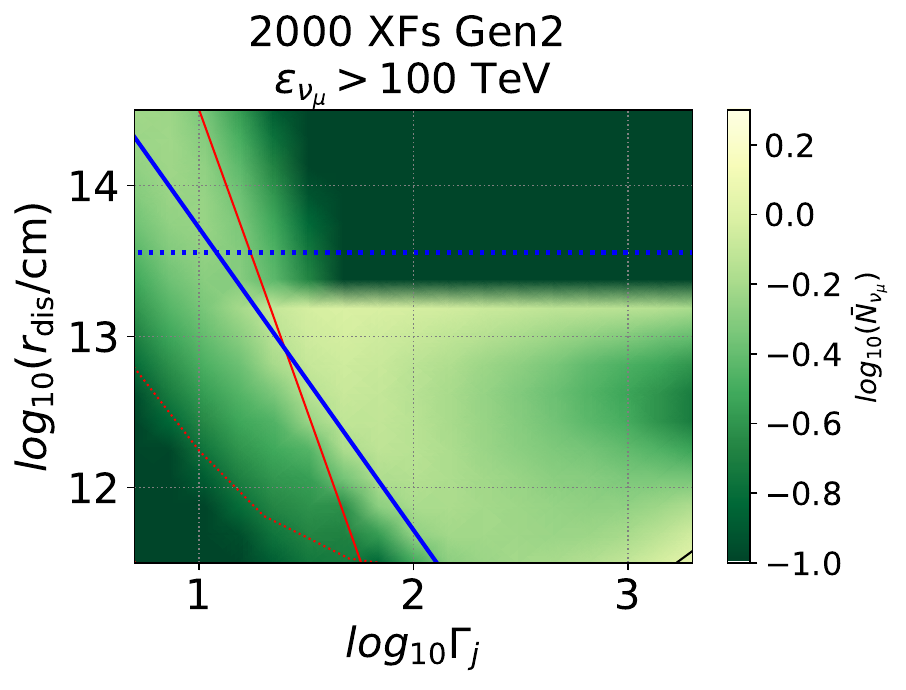}
      \end{minipage}
      
      \begin{minipage}[t]{0.45\hsize}
        \centering
        \includegraphics[keepaspectratio, scale=0.6]{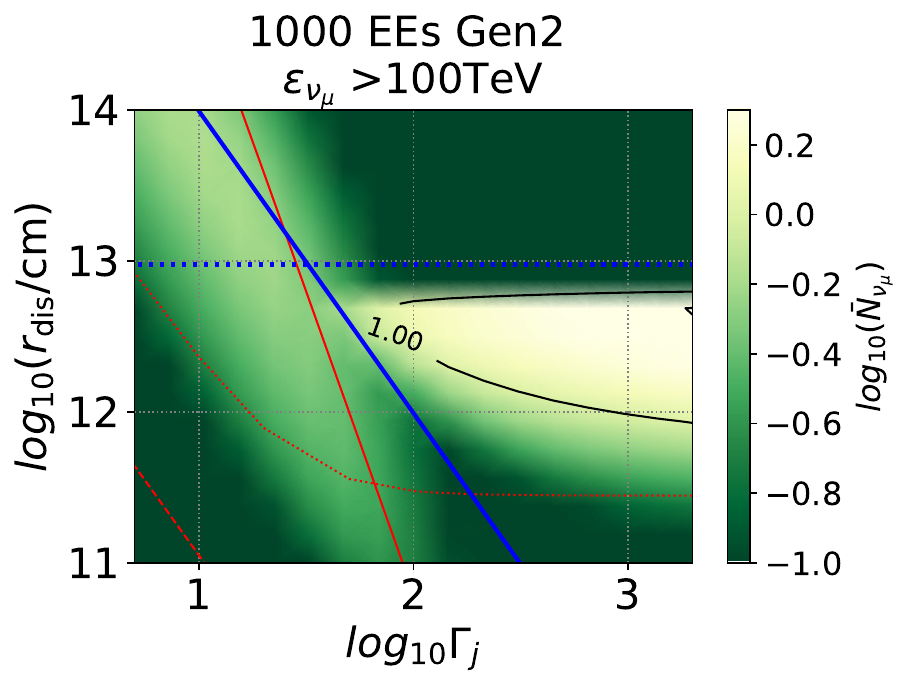}
      \end{minipage}
    \end{tabular}
    \caption{Color maps of the expected number of neutrino detection for 2000 XFs (left) and 1000 EEs (right) with IceCube-Gen2 with threshold at $\varepsilon_{\nu_\mu} = 10$ TeV (top) and at $\varepsilon_{\nu_\mu} = 100$ TeV (bottom). Lines are same as \Fg\ref{fig:number_fid}.
}
    \label{fig:number_fid_cut}
  \end{figure*}

Setting the threshold also cuts off part of the signals. 
The upper two panels in \Fg \ref{fig:number_fid_cut} show $\bar{N}_{\nu_\mu,\rm stacked}$ for XFs ($N_{\rm GRB} = 2000$) and for EEs ($N_{\rm GRB} = 1000$) with $\varepsilon_{\nu_\mu, \rm th} = 10$ TeV. 
They indicate that $\varepsilon_{\nu_\mu, \rm th} = 10$ TeV does not strongly change the results, compared to \Fg \ref{fig:number_fid}. 
However, $\varepsilon_{\nu_\mu, \rm th} = 100$ TeV significantly cuts the signals, as shown by the two lower panels in \Fg \ref{fig:number_fid_cut}. 
This results from the spectral peak at $\varepsilon_{\nu_\mu} \sim 10 - 100$ TeV in most cases. 
This is consistent with the spectrum shown in \Fg \ref{fig:fluence_fid} for our fiducial parameters. Thus, the best strategy to constrain the $\Gamma$-$r_{\rm diss}$ plane is the stacking analysis of neutrinos with a threshold at $\sim 10$ TeV.

\section{Discussion} \label{sec:Dis}
Our results (top panels of \Fg \ref{fig:number_fid}) are consistent with the non-detection of neutrinos in extended time window analysis by IceCube \citep{Abbasi2022grblate} for any ($r_{\rm diss}$, $\Gamma$) in our calculations, assuming $\xi_p =10$.
Thus, with the non-detection, we cannot put a constraint on ($r_{\rm diss}$, $\Gamma$) with 90\%-confidence level for $\xi_p =10$.

In the near future, high-energy neutrino observations could constrain the parameter space based on our results in \Fg \ref{fig:number_fid} and \ref{fig:number_fid_cut}.
In this section we show the prospects of future constraints and their implications on understanding the characteristics of the late-time  jets of GRB.

First, here we consider the operation time required for stacking thousands of late-time emissions. \cite{Abbasi2022grblate} conducted a search for neutrinos in 7.5-year IceCube data associated with 1774 LGRBs and 317 SGRBs listed in the Fermi, Swift, and IPN catalogs \citep{Ajello2019ApJ...878...52A, vonKienlin2020ApJ...893...46V, Lien2016ApJ...829....7L, Hurley2013ApJS..207...39H}.
Extrapolating from this sample, we estimate that it would take approximately 8.5 years (23.6 years) to observe a total of 2000 LGRBs (1000 SGRBs). 
Considering that $\sim$ 30\% of LGRBs have XFs and $\sim$ 50 \% of SGRBs have EEs, it would take $\sim$ 28 years ($\sim$ 47 years) to observe a total of 2000 LGRBs (1000 SGRBs) associated with XFs (EEs).

In the IceCube-Gen2 era, the operation time will be even shorter with future nanosatellite constellation, such as CAMELOT \citep{Werner2018CAMELOT} and HERMES \citep{Fuschino2019HERMES}. 
These facilities will improve the duty cycle and field of view, thus increasing the number of observed GRBs.
With ideal all-sky monitoring of GRBs, it would take $\sim10$ years ($\sim17$ years) to accumulate 2000 XFs (1000 EEs).
Here, we assume that the field of view and the duty cycle of current detectors are $0.7\times 4\pi$ and 0.5, respectively.
%

With respect to the stacking of GRBs, this work ignores the luminosity function and duration distribution of late-time emissions, because they may not significantly affect the results.
For XFs, \cite{Yi2016ApJS..224...20Y} shows the empirical relation $L_{\rm X, iso} \propto t_{\rm dur}^{-1.02}$. The relation indicates that the X-ray fluence ($L_{\rm X, iso} t_{\rm dur}$) is roughly constant for all events. 
Neutrino fluence is roughly proportional to X-ray fluence as long as $f_{p\gamma} \sim 1$, leading to a small dispersion of neutrino fluence.
For EEs, the duration distribution and the empirical relation $L_{\rm X,iso} \propto t_{\rm dur}^{-2.5}$ are shown by \cite{Kisaka2017ApJ...846..142K}. 
The distribution of the duration peaks at $t_{\rm dur} \sim 300\ \rm s$.
This suggests that EEs with $t_{\rm dur} \sim 300\ \rm s$ provide the dominant contribution to the neutrino fluence even if we take into account the luminosity function of EEs. 

For the fiducial case of XFs and LGRBs, the luminosity of the prompt emission is given by $L_{\gamma \rm,pro,iso} \equiv L_{\rm k,pro,iso} f_{\gamma} = 2.1 \times 10^{51} L_{\gamma \rm,iso,48.5} t_{\rm dur, pro,1}(t_{\rm dur}/650\ \mathrm{s})\ \mathrm{erg/s}$. 
Ideally, we should take into account the variation of $L_{\rm k, pro,iso}$ due to $L_{\rm k, pro,iso}\propto E_{\rm coc,i}\propto T_{\rm coc}^4\propto (t^{\prime -1}_{p\gamma})^{4/3}$.
However, it may not significantly affect the result for the following reasons.
Firstly, $f_{p\gamma} \gtrsim 1$ for our fiducial parameters.
Secondly, a lower $L_{\rm \gamma, pro,iso}$ decreases $f_{p\gamma}$, leading to low neutrino fluence, but such low luminous LGRBs are infrequent \citep{Wanderman2010MNRAS.406.1944W}.
Lastly, a higher $L_{\gamma \rm,pro,iso}$ decreases the neutrino cutoff energy $\varepsilon_{\nu,\rm cut}= \varepsilon^\prime_{p, \rm cut}\Gamma/20$, but $\varepsilon_{\nu,\rm cut}\gtrsim 10 {\rm~TeV}$ is satisfied as long as $L_{\gamma \rm,pro,iso} < 10^{53}$ erg/s.
$L_{\gamma \rm,pro,iso} = 10^{53}$ erg/s is beyond the peak of the luminosity distribution and is subdominant for stacking analysis \citep{Wanderman2010MNRAS.406.1944W}.
Thus, the fixed $L_{\rm k,pro,iso}$ is reasonable approximation for the detectability of high-energy neutrinos.
 
Also the variation of $L_{\rm k, pro,iso}$ may not significantly affect the result for EEs in SGRBs.
The reason is the same as for XFs, except for the event rate of the less luminous SGRBs.
For EEs, we use $L_{\gamma \rm,pro,iso} = 3 \times 10^{50}\ L_{\gamma \rm,iso,49} t_{\rm dur, pro,0}(t_{\rm dur}/300\ \mathrm{s})\ \mathrm{erg/s}$.
SGRBs with a lower luminosity are frequently observed \citep{Sakamoto2011ApJS..195....2S,Lien2016ApJ...829....7L,Wanderman2015MNRAS.448.3026W}.
However, even for the lowest luminosity ($L_{\gamma \rm,pro,iso} \sim 3\times 10^{49}$ erg/s), $f_{p\gamma} \sim 1$ is satisfied for a wide range because $t^{\prime-1}_{p\gamma}$ lowered only by $10^{-3/4} \sim 1/6$.

Here, we did not consider the variation of the jet opening angle ($\theta_j$) for the late-time jet, although they could be less collimated \citep{Lu2023MNRAS.522.5848L}.
For a wider $\theta_j$, the suppression by Thomson scattering is enhanced, (the value of $f_{\tau_j}$ is lowered).
However, overall the general trend in \Fg \ref{fig:number_fid} and \ref{fig:number_fid_cut} should be unchanged. 

The correlation between late-time emissions and neutrino data taken by IceCube-Gen2 in the near future supports $\Gamma > 100$ and $r_{\rm diss} < 10^{13.5}$ cm (XFs) or $10^{12}\ {\rm cm} < r_{\rm diss} < 10^{13}$ cm (EEs).
In this region on the $\Gamma$-$r_{\rm diss}$ plane, the radial optical depth of the jet, $\tau_{\rm j,rad} = \tau_{j}/(\theta_j \Gamma)$, is less than unity. 
This condition requires that the dissipation process occurs in the optically thin region and disfavors the dissipative photosphere model \citep{Rees2005ApJ...628..847R}.
Furthermore, the relatively small $r_{\rm diss}$ indicates that the canonical internal shock model is more likely than the ICMART model \citep{Zhang2011ApJ...726...90Z}.

Given that internal shocks drive the dissipation, the variability timescale is expected to be $\delta t \sim r_{\rm diss}/(2c\Gamma^2) \leq 10^{-1}$ s in the case of the neutrino detection. 
This timescale is much shorter than the duration of late-time emissions, but it can be consistent with X-ray observations.
Typically, XRT count rate is $\sim100$ photons per second for XFs and EEs.
We cannot obtain a sophisticated light curve with short time bins to analyze variability in time shorter than 0.1 s.
Quick follow-up observations with larger telescopes, such as NICER \citep{Gendreau2016NICER}, are helpful for understanding the short-time variability in late-time emission.

Estimation of the variability timescale suggests that the central engine could also be variable in a shorter time period than $\sim$ 0.1 s if neutrinos are detected.
This can potentially constrain the physical origin of the late activity. 
One candidate for the central engine is the Blandford-Znajeck (BZ) process with the central brack hole.
Recent GRMHD simulations shows that the accretion rate and jet efficiency are variable in 1000 $r_g/c$ ($r_g$ is the gravitational radius) for the central black hole with Magnetically Arrested Disk (MAD) \citep[e.g.][]{Tchekhovskoy2011MNRAS.418L..79T, Takahashi2016GRRMHD,Ripperda2022ApJ...924L..32R,Utsumi2022ApJ...935...26U}.
For GRBs, the mass of the central black hole should be a few $M_\odot$, indicating that variability timescale of the engine is $\sim0.03$ s.
This is a possible explanation for the short variability of the central engine.
Another candidate for the central engine is the magnetar spin down.
\cite{Das2024NS_GRMHD} shows that the jet power develops relatively smoothly in time with the neutron star engine.
Hence, future neutrino detections would favor the BZ process as the powering mechanism in the late-time jets.

In addition, neutrino detection in the future would confirm the baryonic jet model for the late-time jet.
Some studies have suggested that wind from the newborn magnetar with leptonic process as a possible scenario for the origin of the late-time emission \citep{Dai1998magnetar}.
We can disfavor the leptonic scenario in case of a neutrino detection.
On the other hand, the baryonic jet model requires a baryon loading process prior to dissipation.
For prompt jets, the candidate baryon injection process is neutron diffusion from neutron-rich material \citep{Belovobodov2003neutron,Levinson2003neutron}, which requires huge mass accretion rate to the remnant $\sim 1\ M_\odot /\rm s$ \citep{Kohri2005ndaf}.
Such a high accretion rate is unlikely to occur in a late time, so we need an alternative baryon injection processes in the case of neutrino detection.

Note that the above discussions are based on a value of the cosmic ray loading factor as $\xi_p = 10$.
$\xi_p$ is less constrained for the late-time jet.
Furthermore, its value is directly proportional to the neutrino fluence, which leads to more ambiguity in the parameter constraints.
The non-detection of neutrinos favors the jet with a small fraction of baryons.

Another procedure to constrain $\xi_p$ is to calculate the electromagnetic (EM) emission caused by photohadronic interactions.
The $p\gamma$ process produces neutral pions ($\pi^0$) and charged pions ($\pi^{\pm}$).
The $\pi^0$ decays into two gamma-rays, while the $\pi^{\pm}$ decays into neutrinos, electrons, and positrons.
The energy budget of gamma-rays from $\pi^0$ is comparable to that of neutrinos because photo-hadronic interactions produce a similar amount of $\pi^{0}$ and $\pi^{\pm}$.
For the fiducial case, cosmic rays contribute $f_\gamma \xi_p = 0.3$ as a fraction of the total jet power.
Approximately one-third\footnote{This is determined by ${\rm ln}(\varepsilon^{\prime}_{p,\rm cut}/{\varepsilon}^{\prime}_{p,\rm thrsh})/{\rm ln}(\varepsilon^{\prime}_{p,\rm cut}/\varepsilon^{\prime}_{p,\rm min})\sim1/3$ for our fiducial parameters of XFs.} of them lose their energy through the $p\gamma$ process.
Half of the lost energy is converted into gamma-rays (via $\pi^0$) or neutrinos (via $\pi^\pm$).
Thus, the gamma-rays from $\pi^0$ receive 5\% of the total jet power, which is comparable to internal photons ($f_\gamma = 3\%$).
The gamma-rays should be observed in the $<$ MeV-GeV range due to the EM cascade.
The secondary electrons produced by the cascade could emit X-rays comparable to the late-time emission.
By comparing observational data with theoretical predictions, we can obtain further constraints on the parameter space. 
A consistent calculation of the interactions between internal photons, cocoon photons, and gamma-rays from $\pi^0$ is required.

A potential signature could be detected in GeV gamma-rays observed by Fermi/LAT \citep{Atwood2009LAT}, but it is challenging to distinguish between hadronic and leptonic emissions in the GeV range. 
Cocoon photons can be upscattered by electrons accelerated in the dissipation of the late-time jet \citep{Kimura2019ApJ...887L..16K}, providing a possible explanation for GeV emissions from GRB 211211A \citep{Mei2022Natur.612..236M}. 
The gamma-ray flux by the leptonic scenario may be comparable to that of hadronic scenarios.
It is beyond the scope of this paper to calculate leptonic and hadronic emissions, and to suggest how to distinguish them.

Hadronic emissions are expected to exhibit flat energy fluence from 1 eV to 1 GeV 1000 s after the burst.
Multi-wavelength rapid follow-up and wide field surveys are helpful to probe late-time emissions.
For example, observations by ULTRASAT \citep{Shvartzvald2024ULTRASAT}, MAXI \citep{Matsuoka2009PASJ...61..999M}, Einstein Probe \citep{Yuan2015EinsteinProbe}, HiZ-GUMDAM \citep{Yonetoku2014HiZ}, THESEUS \citep{Amati2018THESEUS}, eASTROGAM \citep{DeAngelis2017ExA....44...25D}, GRAMS \citep{Aramaki2020APh...114..107A}, and AMEGO-X \citep{Caputo2022arXiv220804990C} would constrain the physical parameters of late-time emissions.
Rapid optical follow-up observations by ground-based telescopes \citep{Becerra2023opticalGRB} should also be important.
Such a multi-messenger approach will be helpful in obtaining physical parameters of the late-time jet.

\section{Conclusions} \label{sec:Conc}
We presented a calculation of the neutrino emissions associated with late-time emissions (XFs and EEs) in GRBs, taking into account the photon field from the jet-heated cocoon.
Our spectra shown in \Fg \ref{fig:fluence_fid} indicate that cocoon photons significantly enhance the neutrino emission in 10 TeV to 1 PeV, which is the sensitive energy range of IceCube and IceCube-Gen2.

We also calculated the detectability of the neutrino by these facilities.
The color maps in \Fg \ref{fig:number_fid} shows the expected number of neutrinos from a stacking analysis of thousands of late-time emissions, including cosmological ones.
We found that neutrinos can be detectable once we have data for 2000 XFs or 1000 EEs with IceCube-Gen2, which would be achievable in $\sim$ 10-20 years.
We also confirmed that the detectability of neutrinos depends weakly on the Lorentz factor of the jet if cocoon photons can diffuse into the dissipation region 
as previously shown in \cite{Matsui2023ApJ...950..190M}.

Noise from atmospheric and astrophysical background event likely contaminate (30\% chance) the stacking analysis of thousands of GRBs with a $\sim$ 300 s time window without any threshold.
To reduce the background rate below 1\%, using an energy threshold of $> 10$ TeV is required. 
With \Fg \ref{fig:number_fid} (without neutrino energy threshold) and \Fg \ref{fig:number_fid_cut} (with neutrino energy threshold), we showed that using a neutrino energy threshold (at 10 TeV) reduces the probability of background contamination to 1 \%, and therefore is the most effective way to detect neutrino signals.

With our analytic modeling and an energy threshold at 10 TeV, future observations by IceCube-Gen2 will significantly and sufficiently constrain the Lorentz factor, dissipation radius, and cosmic-ray loading factor of the late-time jets of GRBs.
Hence, future high-energy neutrinos observations, in coordination with EM observations of GRBs, will be a powerful tool to answer some of the most fundamental questions regarding the launch and emission of GRB jets.

\begin{acknowledgments} 
The authors thank Kunihito Ioka, Kazumi Kashiyama, Katsuaki Asano, and Kenji Toma for meaningful discussions. 
This work is supported by Graduate Program on Physics for the Universe (GP-PU), Tohoku University JST SPRING, Grant Number JPMJSP2114(R.M.), and JSPS KAKENHI Nos. 22K14028, 21H04487, 23H04899 (S.S.K.), and 23K19059 (H.H.). 
S.S.K. acknowledges the support by the Tohoku Initiative for Fostering Global Researchers for Interdisciplinary Sciences (TI-FRIS) of MEXT's Strategic Professional Development Program for Young Researchers.
\end{acknowledgments}

\appendix
\section{The case of Black-body photons from the cocoon}
\label{app:diffusion}
Here, we consider a more optimistic scenario than the fiducial one, where the suppression factor $f_{\rm \tau_{coc},skin}$ is set to 1.
We do not consider the cocoon heating by the late-time jet itself for simplicity.
The shock between the cocoon and the late-time jet \citep{Hamidani2023latetime} or the magnetic instability in the limb of the late-time jet can heat cocoon.
This heating might enhance internal energy transportation at the surface of the cocoon.
By this heating, the suppression factor $f_{\tau_{\rm coc},\rm skin}$ can be unity, leading to more efficient neutrino productions.

\begin{figure*}\hspace{-1cm}
    \begin{tabular}{cc}
      \begin{minipage}[t]{0.5\hsize}
        \centering
        \includegraphics[keepaspectratio, scale=0.4]{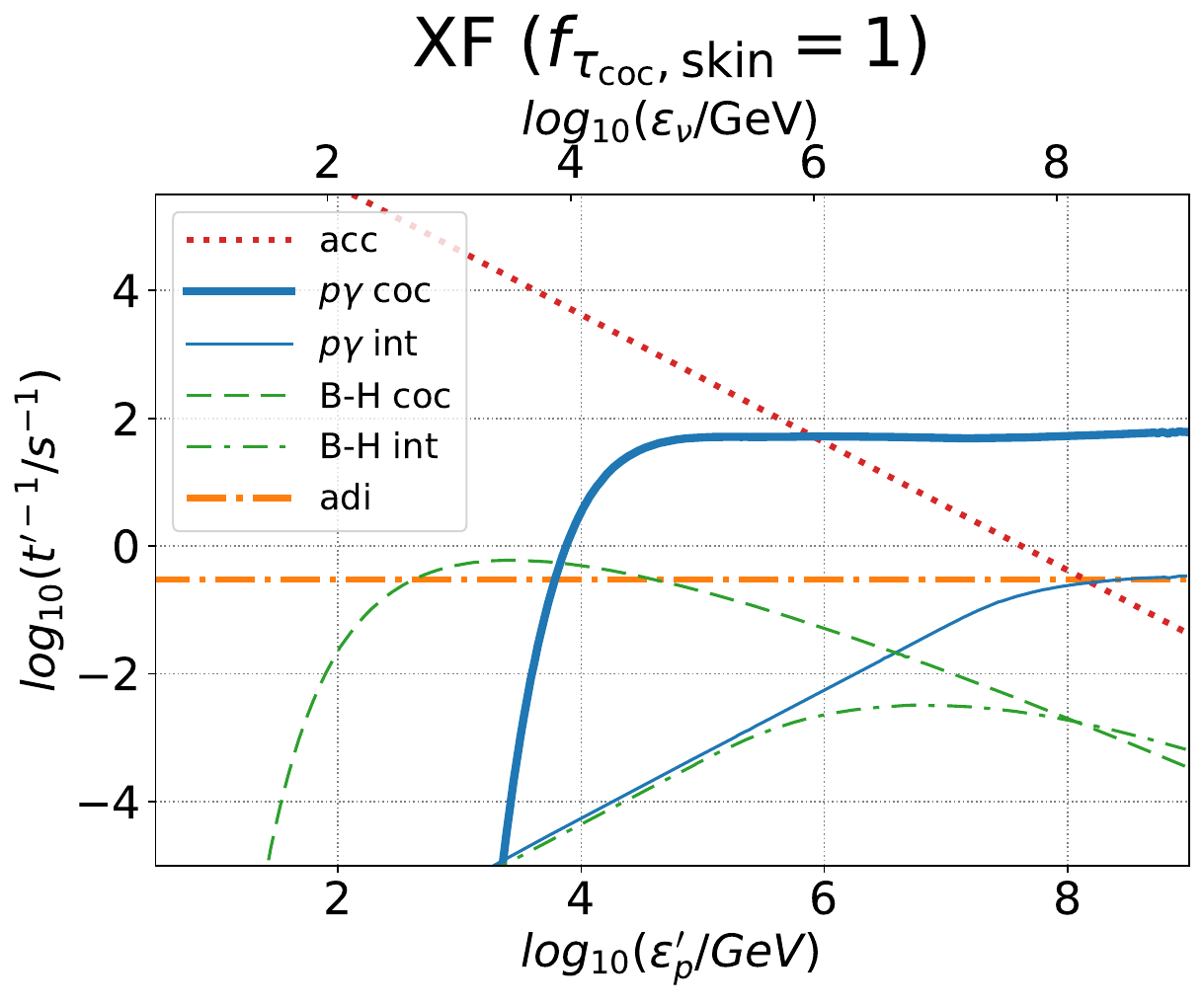}
      \end{minipage}
      
      \begin{minipage}[t]{0.5\hsize}
        \centering
        \includegraphics[keepaspectratio, scale=0.4]{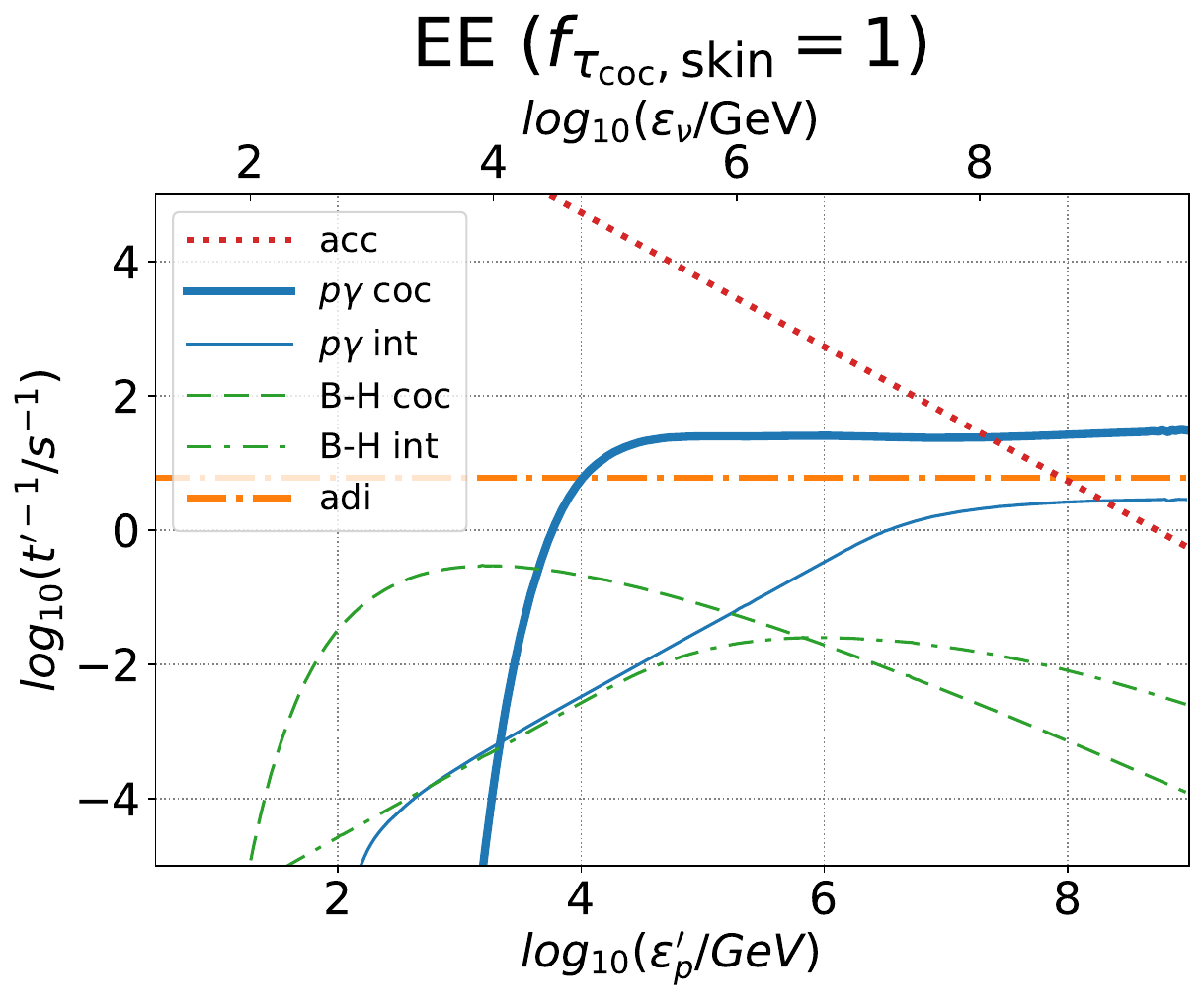}
      \end{minipage}
      
    \end{tabular}
    \caption{The cooling and acceleration rate for our fiducial parameters of XF (left) and EE (right) for $f_{\tau_{\rm coc},\rm skin} =1$. Lines are same as \Fg\ref{fig:time_fid}.}
    \label{fig:time_BB}
  \end{figure*}

We calculate the neutrino fluence and its detectability for $f_{\tau_{\rm coc},\rm skin} =1$ as an optimistic case.
The cooling rate for $f_{\tau_{\rm coc},\rm skin} =1$ with our fiducial parameters is shown in \Fg \ref{fig:time_BB}, and its neutrino fluence is represented by the orange lines in \Fg \ref{fig:fluence_fid}.
Their cutoff energies for neutrinos are slightly lower than those in the fiducial case ($f_{\tau_{\rm coc},\rm skin} \simeq 0.93$ for XF and $f_{\tau_{\rm coc},\rm skin} \simeq 0.19$ for EE).
This is because the $p\gamma$ cooling is slightly higher without the suppression by $f_{\tau_{\rm coc},\rm skin}$.
Nevertheless, the resulting spectra are similar to those for our fiducial models. 

\begin{figure*}\hspace{-1.5cm}
    \begin{tabular}{cc}
      \begin{minipage}[t]{0.5\hsize}
        \centering
        \includegraphics[keepaspectratio, scale=0.6]{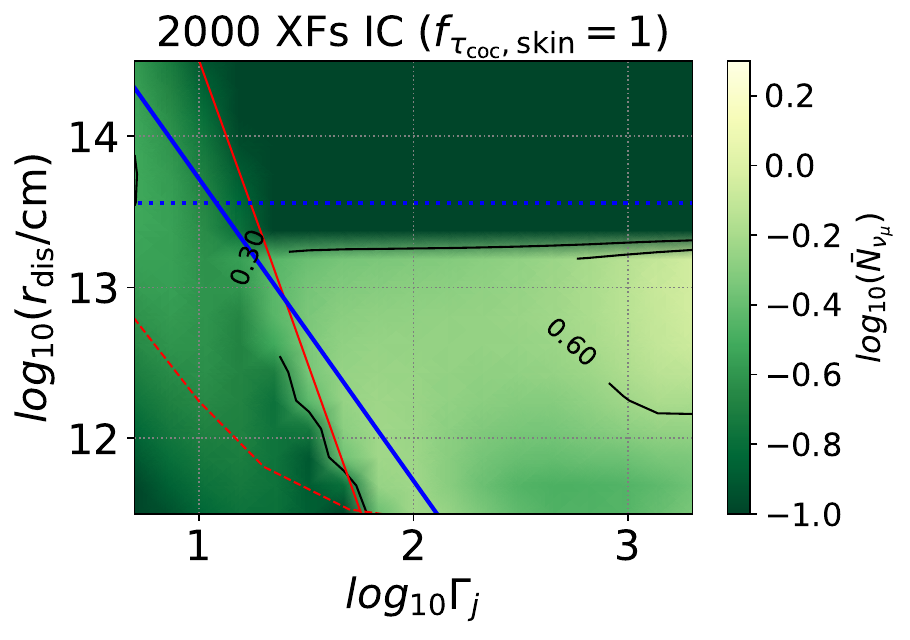}

      \end{minipage} 
      \begin{minipage}[t]{0.45\hsize}
        \centering
        \includegraphics[keepaspectratio, scale=0.6]{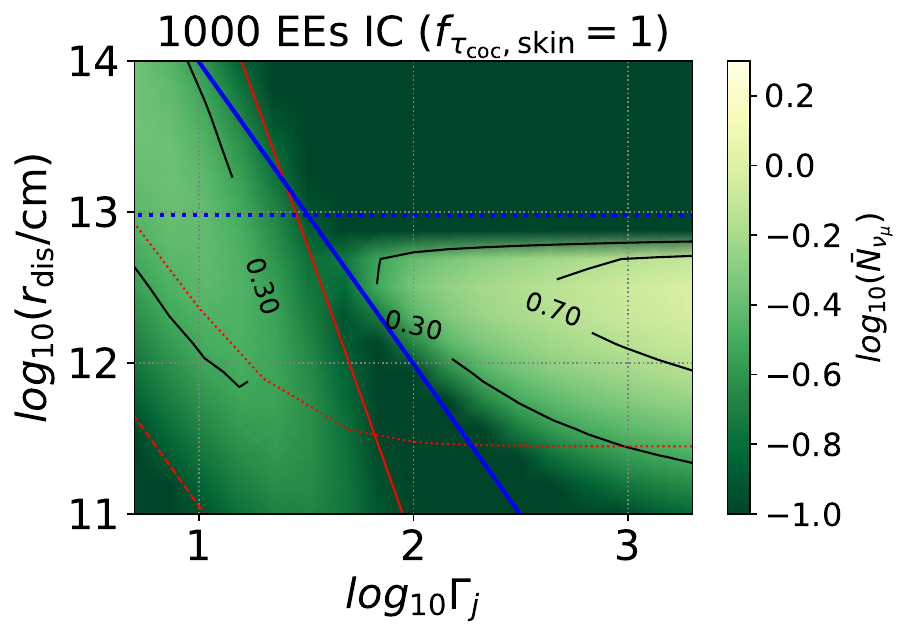}
      \end{minipage}\\
      
      \begin{minipage}[t]{0.5\hsize}
        \centering
        \includegraphics[keepaspectratio, scale=0.6]{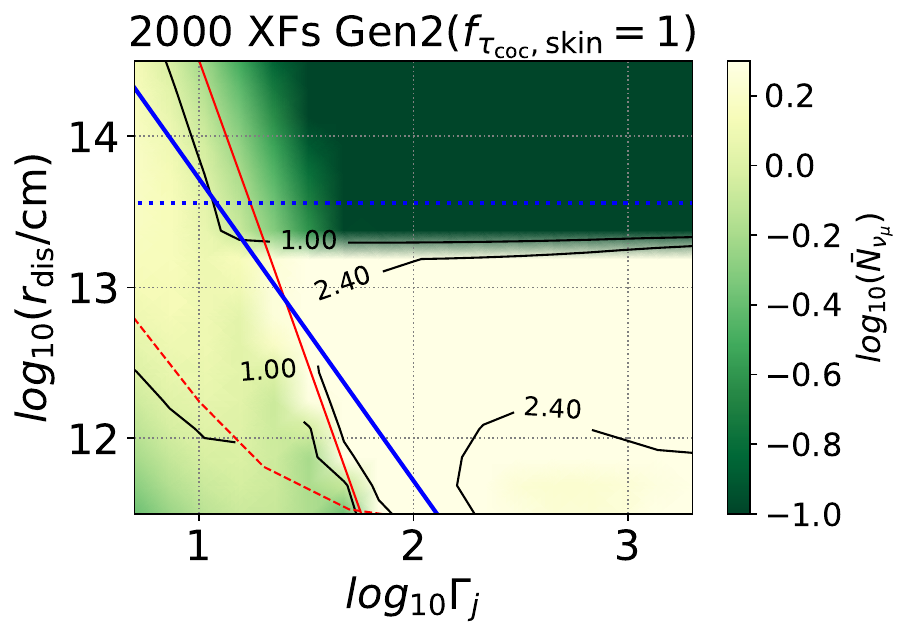}
      \end{minipage}
      
      \begin{minipage}[t]{0.45\hsize}
        \centering
        \includegraphics[keepaspectratio, scale=0.6]{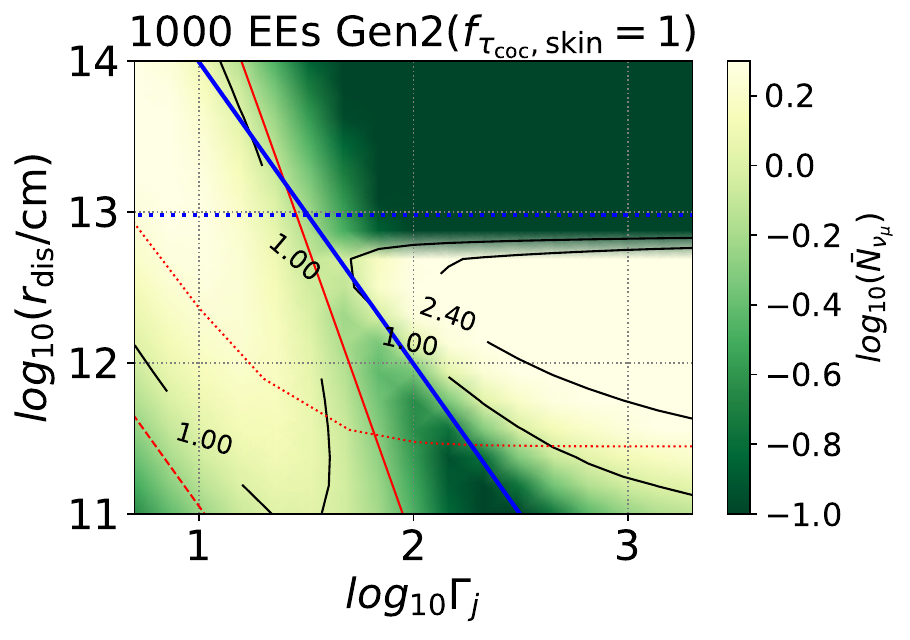}
      \end{minipage}
    \end{tabular}
    \caption{Color maps of the expected number of neutrino detection for 2000 XFs (left) and 1000 EEs (right) with IceCube-Gen2 without the energy threshold for $f_{\tau_{\rm coc},\rm skin} =1$. Lines are same as \Fg\ref{fig:number_fid}.
}
    \label{fig:number_BB}
  \end{figure*}

\begin{figure*}\hspace{-1.5cm}
    \begin{tabular}{cc}
      \begin{minipage}[t]{0.5\hsize}
        \centering
        \includegraphics[keepaspectratio, scale=0.6]{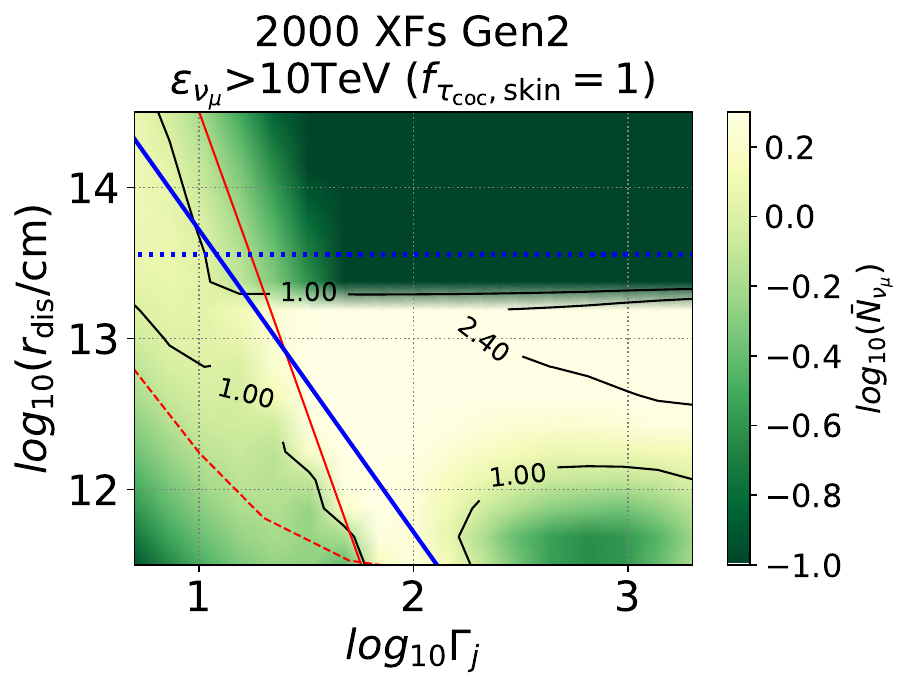}
      \end{minipage} 
      \begin{minipage}[t]{0.45\hsize}
        \centering
        \includegraphics[keepaspectratio, scale=0.6]{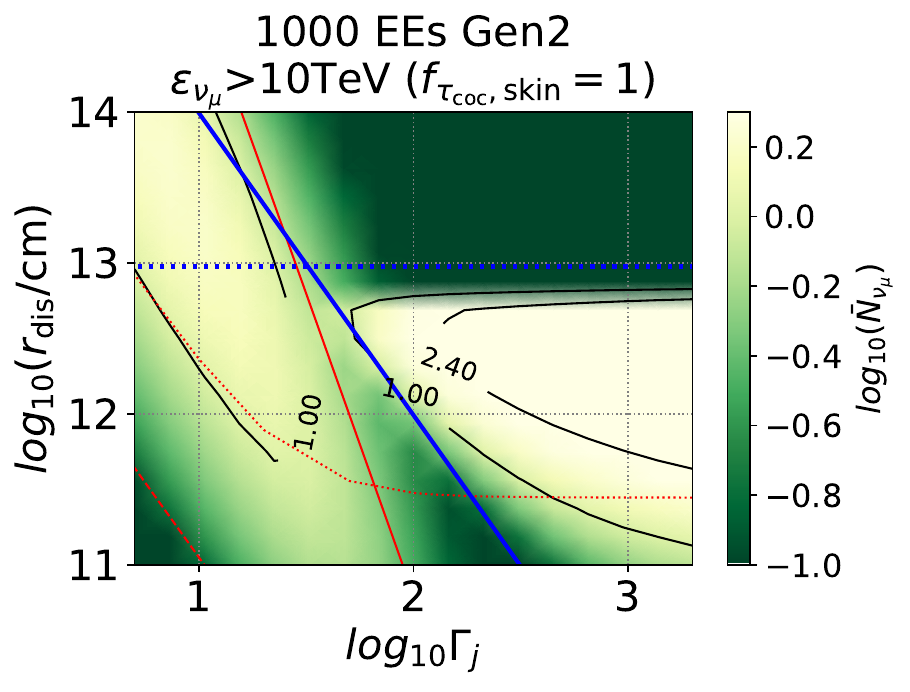}
      \end{minipage}\\
      
      \begin{minipage}[t]{0.5\hsize}
        \centering
        \includegraphics[keepaspectratio, scale=0.6]{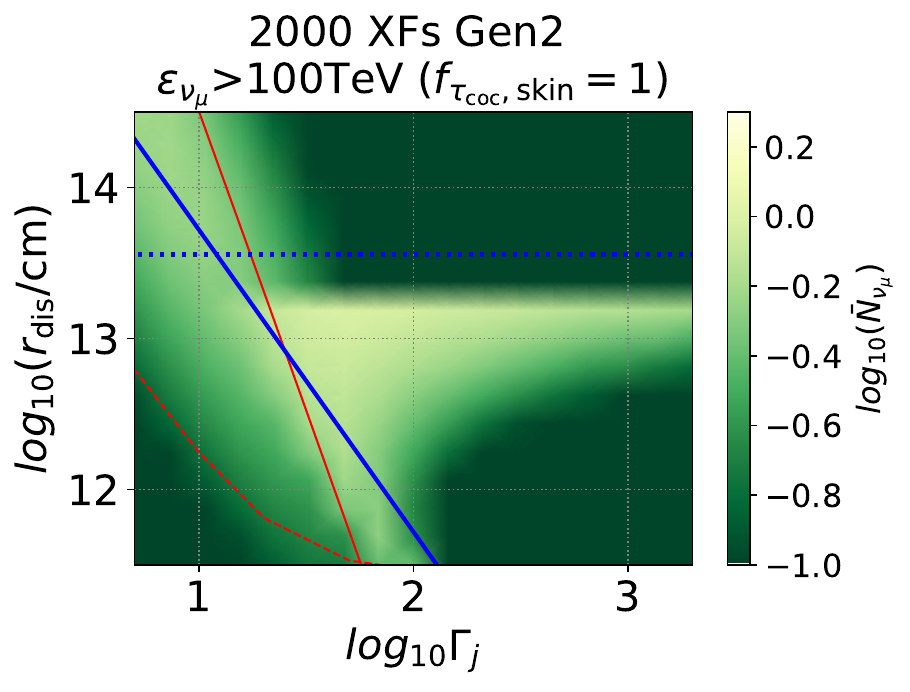}
      \end{minipage}
      
      \begin{minipage}[t]{0.45\hsize}
        \centering
        \includegraphics[keepaspectratio, scale=0.6]{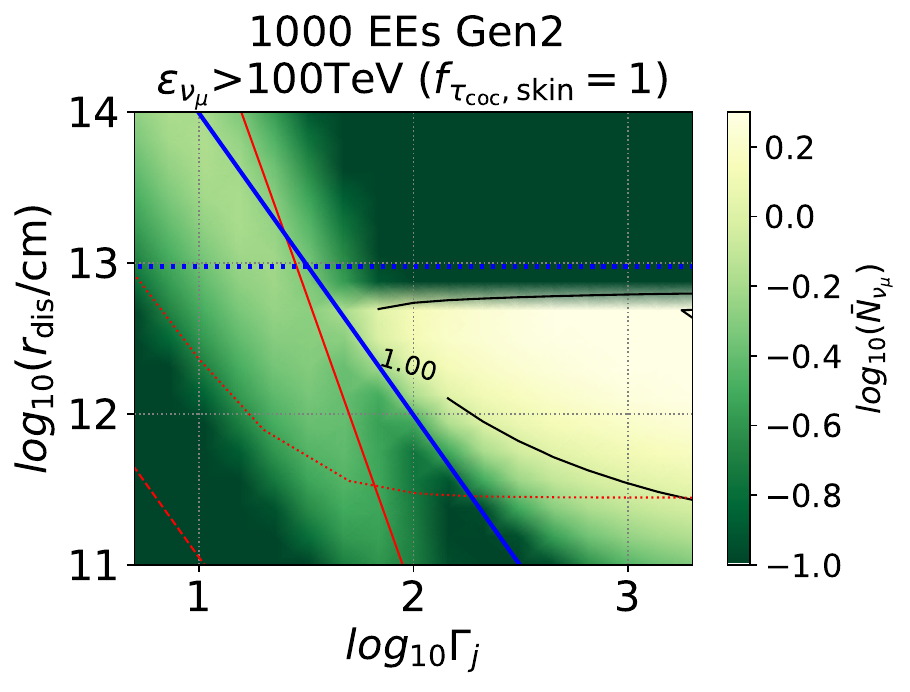}
      \end{minipage}
    \end{tabular}
    \caption{Color maps of the expected number of neutrino detection for 2000 XFs (left) and 1000 EEs (right) with IceCube-Gen2 with threshold at $\varepsilon_{\nu_\mu} = 10$ TeV (top) and at $\varepsilon_{\nu_\mu} = 100$ TeV for $f_{\tau_{\rm coc},\rm skin} =1$. Lines are same as \Fg\ref{fig:number_fid}.
}
    \label{fig:number_BB_cut}
  \end{figure*}

The expected detection numbers in wide range of $r_{\rm diss}$ and $\Gamma$ for $f_{\tau_{\rm coc},\rm skin} =1$ are also shown in \Fg \ref{fig:number_BB} (without the threshold) and \Fg \ref{fig:number_BB_cut} (with the threshold).
For XFs, setting $f_{\tau_{\rm coc},\rm skin} =1 $ negatively affects the expected number of neutrino detection.
For $r_{\rm diss} \sim 10^{12}$ cm and $\Gamma\sim10^3$, $t_{p\gamma}^{\prime-1}$ is so high that the neutrino cutoff energy ($\varepsilon_{\nu,\rm cut}$) is lower than 10 TeV, leading to the low expected number of this region.
For EEs, $f_{\tau_{\rm coc},\rm skin} =1$ increase the detection numbers at $r_{\rm diss} \sim 10^{11}$ cm and $\Gamma>100$ because $t_{p\gamma}^{\prime-1}$ become higher but not too high to let $\varepsilon_{\nu,\rm cut}< 10 $ TeV.

\bibliography{sample631}{}
\bibliographystyle{aasjournal}



\end{document}